\documentclass[preprint,prmaterials,showpacs,twocolumn,showkeys,superscriptaddress,10pt,floatfix]{revtex4-1}

\usepackage{graphicx}
\DeclareGraphicsExtensions{.pdf,.png,.jpg}
\usepackage{epsfig}
\usepackage{transparent}
\usepackage{natbib}
\usepackage[colorlinks=true,linkcolor=blue]{hyperref}
\usepackage{color}
\usepackage[dvipsnames]{xcolor}
\usepackage{soul}
\usepackage[normalem]{ulem} 

\usepackage[toc,page]{appendix}

\usepackage{gensymb}
\usepackage{placeins}
\begin{document}

\title{Magnetic and all-optical switching properties of amorphous Tb$_x$Co$_{100-x}$ alloys}


\author{Agne Ciuciulkaite}
\thanks{Email: agne.ciuciulkaite@physics.uu.se}
\affiliation{Department of Physics and Astronomy, Uppsala University, Box 516, SE-75120 Uppsala, Sweden}

\author{Kshiti Mishra}
\affiliation{Radboud University, Institute for Molecules and Materials, Nijmegen, the Netherlands}

\author{Marcos V. Moro}
\affiliation{Department of Physics and Astronomy, Uppsala University, Box 516, SE-75120 Uppsala, Sweden}

\author{Ioan-Augustin Chioar}
\affiliation{Department of Physics and Astronomy, Uppsala University, Box 516, SE-75120 Uppsala, Sweden}

\author{Richard M. Rowan-Robinson}
\altaffiliation[Present address: ]{Department of Material Science and Engineering, University of Sheffield, Sheffield, United Kingdom}
\affiliation{Department of Physics and Astronomy, Uppsala University, Box 516, SE-75120 Uppsala, Sweden}

\author{Sergii Parchenko}
\affiliation{Swiss Light Source, Paul Scherrer Institute, 5232 Villigen PSI, Switzerland}

\author{Armin Kleibert}
\affiliation{Swiss Light Source, Paul Scherrer Institute, 5232 Villigen PSI, Switzerland}

\author{Bengt Lindgren}
\affiliation{Department of Physics and Astronomy, Uppsala University, Box 516, SE-75120 Uppsala, Sweden}

\author{Gabriella Andersson}
\affiliation{Department of Physics and Astronomy, Uppsala University, Box 516, SE-75120 Uppsala, Sweden}

\author{Carl Davies}
\affiliation{FELIX Laboratory, Radboud University, Nijmegen, the Netherlands}

\author{Alexey Kimel}
\affiliation{Radboud University, Institute for Molecules and Materials, Nijmegen, the Netherlands}

\author{Marco Berritta}
\affiliation{Department of Physics and Astronomy, Uppsala University, Box 516, SE-75120 Uppsala, Sweden}

\author{Peter M.\ Oppeneer}
\affiliation{Department of Physics and Astronomy, Uppsala University, Box 516, SE-75120 Uppsala, Sweden}

\author{Andrei Kirilyuk}
\affiliation{FELIX Laboratory, Radboud University, Nijmegen, the Netherlands}

\author{Vassilios Kapaklis}
\thanks{Email: vassilios.kapaklis@physics.uu.se}
\affiliation{Department of Physics and Astronomy, Uppsala University, Box 516, SE-75120 Uppsala, Sweden}

\date{\today}
\begin{abstract}
        Amorphous Tb$_{x}$Co$_{100-x}$ magnetic alloys exhibit a list of intriguing properties, such as perpendicular magnetic anisotropy, high magneto-optical activity and magnetization switching using ultrashort optical pulses. Varying the Tb:Co ratio in these alloys allows for tuning properties such as the saturation magnetic moment, coercive field and the performance of the light-induced magnetization switching. In this work, we investigate the magnetic, optical and magneto-optical properties of various Tb$_{x}$Co$_{100-x}$ thin film alloy compositions. We report on the effect the choice of different seeding layers has on the structural and magnetic properties of Tb$_{x}$Co$_{100-x}$ layers. We also demonstrate that for a range of alloys, deposited on fused silica substrates, with Tb content of 24-30 at.$\%$,  helicity dependent all-optical switching of magnetization can be achieved, albeit in a multi-shot framework. We explain this property to arise from the helicity-dependent laser induced magnetization on the Co sublattice due to the inverse Faraday effect. Our study provides an insight into material aspects for future potential hybrid magneto-plasmonic TbCo-based architectures. 
        \end{abstract}
        
\maketitle

\section{Introduction}

The development of magnetic memory storage devices was accelerated by advancements in nano-fabrication and material preparation methods. However, for dense and fast magnetic memory devices, optimal materials and nanostructures are essential. Rare earth-transition metal (RE-TM) alloys have become a material class of interest after the first observation of all-optical switching (AOS) of their magnetization, using ultrashort laser pulses, as in the case of ferrimagnetic GdFeCo alloys \cite{Stanciu_AO_circularly_polarized_light, GdFeCo-Stanciu}.
Many studies have been carried out in the area of AOS, ranging from its observation in different classes of magnetic materials \cite{Hadri-review}, such as ferrimagnetic alloys \cite{kirilyuk_laser-induced_2013,cheng_laser_2017}, compensated ferrimagnets \cite{Schubert_AOHDMS_ArtifZeroMomMagn}, multilayers \cite{aviles-felix_single-shot_2020}, and even ferromagnetic films \cite{Lambert1337,John2017} and ferro- and ferrimagnetic multilayers \cite{10.3389/fmats.2016.00008,Alebrand_TbCo_magn_rev,Mangin,Beens2019}, e.g.\ Pt-sandwiched Co films \cite{vomir_single_2017}, and Gd/Co bilayers \cite{Lalieu2017} to the exploration of the parameter space for the observation of AOS. The latter includes material compositions \cite{doi:10.1002/adma.201300176, hadri_domain_size_criterion_PhysRevB.94.064419} and laser pulse parameters, such as fluence \cite{alebrand_subpicosecond_2014}, helicity \cite{Alebrand_interplay, ElHadri2016}, and pulse duration \cite{Medapalli_Multiscale}. A range of both experimental and theoretical works were performed in order to explain the mechanism of AOS: ultrafast heating \cite{ultrafast-heating-ferrimagnet,cornelissen_microscopic_2016}, angular momentum transfer between sublattices \cite{Baryakhtar2013} comparing ultrafast heating and coherent effects \cite{Raposo_2020}, numerical studies investigating different spin-coupling strengths \cite{du_prediction_2017}, mean-field models of relaxation dynamics in ferrimagnets \cite{davies2019blueprint},  and \textit{ab initio} calculations \cite{Berritta2016} combined with atomistic spin-dynamics modeling \cite{Wienholdt2013,John2017}.

A `single-shot' switching behavior has been observed in amorphous GdFeCo \cite{Stanciu_AO_circularly_polarized_light}, Gd/Co bilayers \cite{Lalieu2017}, and Co/Pt layers \cite{vomir_single_2017}, but not in other materials, including other ferrimagnetic alloys. Instead, most of the investigated materials show helicity-dependent all-optical switching (HD-AOS) in response to a train of multiple circularly-polarized laser pulses \cite{Mangin,Lambert1337,John2017}. 
 Though single-shot switching of magnetization as observed in GdFeCo \cite{kirilyuk_laser-induced_2013} is highly desirable for the scheme of magnetic data writing, the low spin-orbit coupling of Gd leads to a lower magnetic anisotropy in this material, which may pose problems for long-term information retention as well as the down-scaling of the memory bit size. In this respect, the high spin-orbit coupling of Tb renders TbCo a promising material, exhibiting larger magneto-crystalline anisotropy than Gd-based alloys, which could be more relevant for HD-AOS spintronic-based applications, if optimized for its efficiency  \cite{Modeling-of-TbCo-thermallyInducedAOS}. 
Recently, a single-shot helicity-independent all-optical switching was demonstrated in Tb/Co multilayers, coupled to CoFeB electrodes by \citet{aviles-felix_single-shot_2020}. To the best of our knowledge, this is the first and so far the only observation of single-shot induced magnetization switching in a Tb-Co system. However, no helicity-dependence was demonstrated. Also, it is not yet clearly understood how similar or distinct the dynamics in multilayers
and alloys are, or why the multilayers show AOS.
The magnetic properties of amorphous ferrimagnetic RE-TM Tb$_{x}$Co$_{100-x}$ alloys are highly tunable. 
It has been theoretically predicted that Tb${_x}$Co$_{100-x}$ could exhibit (due to laser heating) thermally-induced  magnetization switching over a wide range of atomic compositions, thus attracting attention as a suitable material for the preparation of the building blocks of non-volatile memory elements \cite{Modeling-of-TbCo-thermallyInducedAOS}. TbCo alloys also exhibit a favorable perpendicular magnetic anisotropy \cite{yoshino_perpendicular_1984, Frisk_2015,M-MO-prop-RE/TM-alloys-1989}, which makes them attractive for densely packed magnetic memory elements. It was recently shown by \citet{Rich_Truncated_nanocones}, that even after a lithographic process to fabricate nanosized truncated Au/Tb$_{18}$Co$_{82}$ nanocones, the out-of-plane magnetic anisotropy is preserved in the magnetic tip of the nanocone. The Tb and Co sublattices interact antiferromagnetically, but their magnetic moments are uncompensated, giving rise to a non-zero net magnetic moment. Starting from a TM-magnetic moment dominated condition  at a given temperature, reduction of the Co content in the Tb$_{x}$Co$_{100-x}$ alloy increases the coercive field, $\mathrm{\mu_0H_c}$, and decreases the saturation moment until no net magnetic moment exists. This point is called the magnetization compensation point $x\mathrm{_{comp}}$, at which the ferrimagnet resembles an antiferromagnet. Upon passing through the compensation point to a RE-magnetic moment dominated composition region, the net magnetic moment of the alloy begins to increase again. Alternatively, an alloy with a given composition can become a compensated ferrimagnet upon adjusting the temperature to the corresponding magnetization compensation temperature $T\mathrm{_{comp}}$ \cite{M-MO-prop-RE/TM-alloys-1989}. 
In this work we present results of a study of the magnetic properties and HD-AOS behavior of amorphous RE-TM Tb$_{x}$Co$_{100-x}$ alloys. We investigate the effect of the atomic ratio of Tb:Co, the thickness of the films and the underlying non-magnetic buffer layers on the structural and magnetic properties of the alloys. We also explore the influence of the laser pulse characteristics, such as fluence, number of shots and sweeping speed, on the AOS performance Tb$_{x}$Co$_{100-x}$ layer. We demonstrate an increased range of the compositions, namely, from 24 to up to 30 at. \% of Tb, for which HD-AOS switching occurs with a laser pulse of 90 fs and 240 fs, depending on the measurement (multi-shot or sweeping beam, respectively), which are, a broader range, and a smaller pulse width, than e.g. reported by \citet{Alebrand_TbCo_magn_rev}. 
In the work by \citet{hadri_domain_size_criterion_PhysRevB.94.064419}, the domain size with the respect to the laser spot size was used as a criterion for the observation of AOS, however, in our study we show, that it is not the only criterion and other parameters, such as the Tb:Co ratio and laser parameters, allow for the observation of HD-AOS irrespective of the domain size. The mechanism for HD-AOS in Tb${_x}$Co$_{100-x}$ samples investigated here is not a single-shot helicity-independent mechanism like in GdFeCo, but rather a cumulative multi-shot mechanism similar to that observed in FePt and Co/Pt multilayers \cite{Mangin,Lambert1337,ElHadri2016}. To explain the origin of the HD-AOS, we further use \textit{ab initio} calculations to compute the local magnetization that can be coherently induced by circularly polarized light through the inverse Faraday effect \cite{Vanderziel1965}. Lastly, an aim of this investigation is to establish a suitable parameter space for the observation of HD-AOS in Tb$_{x}$Co$_{100-x}$ alloys, using a minimal number of laser shots.

\section{Materials and Methods\label{Section:Mat_Meth}}

Amorphous Tb$_{x}$Co$_{100-x}$ alloys were prepared in an ultrahigh vacuum DC magnetron co-sputtering system, from elemental Tb and Co targets (99.99 \% purity) and using Ar as a sputtering gas. The sample holder was rotating during deposition in order to ensure film thickness and composition uniformity. The following stacking structures were fabricated: substrate/spacer/Tb$_{x}$Co$_{100-x}$/capping, where substrate was fused silica, spacer was either Al$_{80}$Zr$_{20}$, Al$_2$O$_3$ or Au/Al$_2$O$_3$ or no spacer was used at all, and capping layer was Al$_{80}$Zr$_{20}$, Al$_2$O$_3$ for respective spacer layers, or Al$_2$O$_3$ capping was used for Au/Al$_2$O$_3$ spacers and for the samples without the spacer layers. TbCo films were then characterized by Rutherford backscattering spectrometry (RBS) and particle induced X-ray emission (PIXE), and by X-ray scattering techniques to be able to probe the conditions for AOS observation for a well-defined composition and atomic structure of the fabricated Tb$_{x}$Co$_{100-x}$ layers. The elemental composition of the samples was investigated by ion beam analysis at the Tandem Laboratory of Uppsala University, using the 5-MV NEC-5SDH-2 tandem accelerator~\cite{Tandem_2019}. RBS and PIXE were employed using 2.0 MeV He$^+$ as a  primary beam. 
X-ray reflectivity (XRR) measurements were performed in order to determine layer thickness as well as roughness. In the data presented in the figures we refer to the nominal, i.e.\ as calibrated, thickness of the films. The XRR determined thickness was 2 - 4 nm smaller than the nominal one (see Supplementary information). The roughness of the layers as determined from fitting of XRR curves using \textit{GenX} \cite{GenX}, was around $5.9$ {\AA} for the samples deposited directly on the fused silica substrates. More values of the root mean square roughness of TbCo layers with varying Tb:Co ratios and buffer layers are provided in the Supplementary Information, Table I. The atomic structure of the films was investigated using grazing incidence X-ray diffraction (GIXRD) measurements in order to determine whether the layers are amorphous or crystalline. More details about the sample preparation and characterization can be found in the Supplementary Information and in Refs.\  \cite{Ciuciulkaite1360280,MORO2019137416}.

We further performed ellipsometry measurements to determine the optical constants, i.e.\ the refractive index $n$ and the extinction coefficient $k$, within the visible and near-infrared range, for various concentrations of Tb$_{x}$Co$_{100-x}$. We measured $20 - 40$ nm thick SiO$_{2}$/Tb$_{x}$Co$_{100-x}$/Al$_2$O$_3$ films with Tb content in the range of $18.4 - 30$ at.$\%$ prepared onto fused silica substrates. For each sample, measurements were taken in the $400 - 1600$ nm spectral range using different incidence angles in the range from $45\degree$ to $75\degree$ in the optical convention, i.e.\ from the sample normal. Finally, the fitting of the ellipsometric data and the extraction of the wavelength dependence of $n$ and $k$ was performed using the $GenX$ software \cite{GenX,genx_source}, specifically adapting it for the visible and near-infrared range and considering a stack layer model with thicknesses determined through XRR measurements and \textit{GenX} fitting of the resulting reflectograms. More details about the $n$ and $k$ extraction are provided in the Supplementary Information.

Static magnetic and magneto-optical properties of the Tb$_{x}$Co$_{100-x}$ films were analyzed employing magneto-optical Kerr effect (MOKE). Magnetization loops were measured using a polar MOKE (PMOKE) setup in reflection geometry with an incident light of 530 nm in wavelength. A maximum out-of-plane static magnetic field of 900 mT was accessible. The remanent magnetization state of Tb$_{x}$Co$_{100-x}$ films was imaged using magneto-optical Kerr microscopy. We used a Kerr microscope with a white light source and applied the magnetic field perpendicular to the sample plane. In order to image the remanent magnetization state, the samples were first demagnetized in a time-dependent magnetic field of decaying amplitude. The obtained micrographs were then analyzed using a pair correlation function (PCF). Element-specific domain imaging combining X-ray photoemission electron microscopy (XPEEM) with the X-ray magnetic circular dichroism (XMCD) effect at the Co $L_3$ and the Tb $M_5$ edges \cite{le_guyader_studying_2012}, was performed at the Surface/Interface: Microscopy (SIM) beamline of the Swiss Light Source. XPEEM was further used to record X-ray absorption (XA) spectra for characterization of the chemical state.

In order to investigate HD-AOS of magnetization in the layer of Tb$_{x}$Co$_{100-x}$, static magneto-optical imaging was performed on SiO$_{2}$/Tb$_{x}$Co$_{100-x}$/Al$_2$O$_3$ samples (deposited directly on fused silica substrates and capped with Al$_2$O$_3$). The samples were illuminated at normal incidence using linearly-polarized white light from an LED source. Upon transmitting through the sample, the light was collected by a 20$\times$ objective lens, passed through a crossed analyzer to obtain magneto-optical contrast, and eventually detected by a CCD camera. We studied the magnetization switching behavior as a function of sample composition and thickness for two types of ultrafast stimuli: (a) keeping the sample stationary and varying the number of laser pulses (`shots') incident on the sample, and (b) sweeping the sample under a train of ultrashort laser pulses. To achieve HD-AOS, pump pulses were generated from a Ti:Sapphire amplified laser system with a 1 kHz repetition rate and a central wavelength of 800 nm. For the multi-shot measurements, the pulse width at the sample was 90 fs and incident at 15$\degree$ to the sample normal. For the sweeping beam measurements, the pump had a pulse width of about 240 fs at the sample position and was incident at 10$\degree$ to the sample normal. A translation stage was used to control the sweep speed for the sweeping beam measurements. 

To investigate the origin of the AOS we performed \textit{ab initio} calculations of the inverse Faraday effect (IFE) \cite{Vanderziel1965} that quantifies the amount of magnetization induced in the material by circularly polarized laser light. To this end we used the non-linear response theory formulation \cite{Battiato2014,Berritta2016}, which gives the induced magnetization $\delta M^{\rm IFE}$ per simulation-cell volume as
\begin{equation}
    \delta M^{\rm IFE} = K^{\rm IFE} (\omega ) \, I /c ,
\end{equation}
where $K^{\rm IFE} (\omega)$ is the frequency-dependent IFE constant, $I$ is the intensity of the laser, and $c$ is the velocity of light. The IFE constant was computed separately for the Tb and Co atoms, assuming here the ordered TbCo$_2$ Laves compound as model system. 

\section{Results}
\subsection{Elemental and structural characterization}

\begin{figure}[!tbp]
  \centering
    \includegraphics[width=\columnwidth]{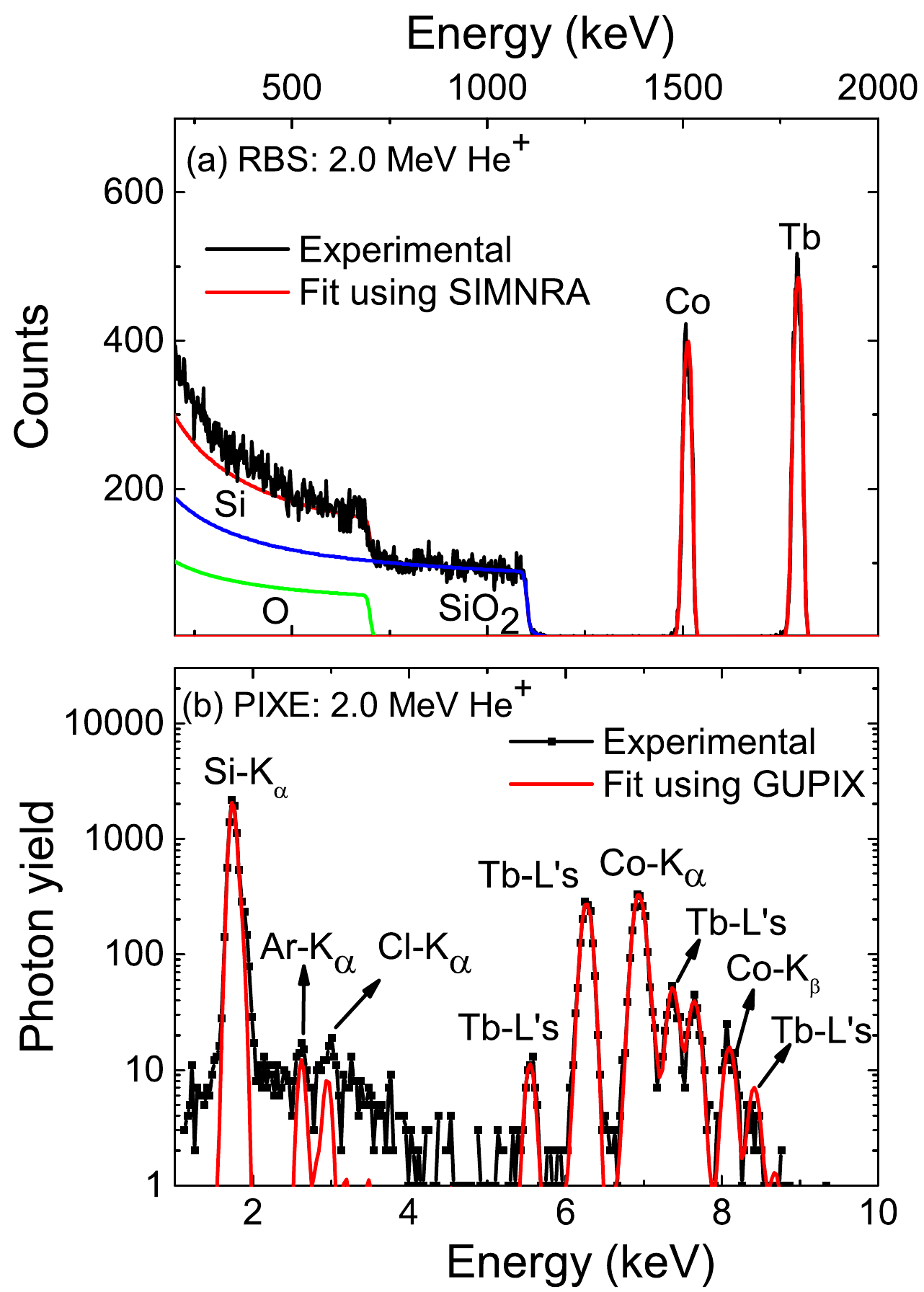}
  \caption{(a) Experimental RBS spectrum recorded for the 2 MeV He$^+$ primary ions scattered from the $\mathrm{Tb_{18}Co_{82}}$ sample prepared on a fused silica substrate (black solid line). The red solid line represents the best fit provided by the SIMNRA code \cite{Mayer_SIMNRA7-1}. The green and blue lines represent the contributions from the O and Si respectively, which arise from the fused silica substrate. (b) The PIXE spectrum (black line with symbols) from the same sample in (a), which is obtained simultaneously with the RBS spectra. The fit using the GUPIX code \cite{Campbell-3} is also shown for comparison (red solid line).
    \label{Fig:RBS+PIXE}
}
\end{figure}

A typical RBS spectrum from $\mathrm{Tb_{18}Co_{82}}$ sample studied in this work is shown in Fig.\ \ref{Fig:RBS+PIXE}(a). The statistical uncertainty of the Tb:Co ratio was found to be $\approx$ 2$\%$, whereas systematic uncertainties were estimated to be lower than 1$\%$. The elemental composition was determined for samples deposited under different deposition conditions in order to obtain the desired Tb:Co ratios. In Fig.\ \ref{Fig:RBS+PIXE}(b), the PIXE spectrum of the same sample is shown. As can be seen in the PIXE fit, trace contaminations of Cl and Ar were found in the sample ($\approx$ 0.2 at.$\%$), while no evidence of heavy trace elements (Z $>$11) was detectable in the film (quantification limit $>$ 0.1 at.$\%$ for the present measurements). The Ar trace contamination found in the films was incorporated during the sample preparation since Ar was used as sputtering gas.

\begin{figure}
    \centering
    \includegraphics[width=\columnwidth]{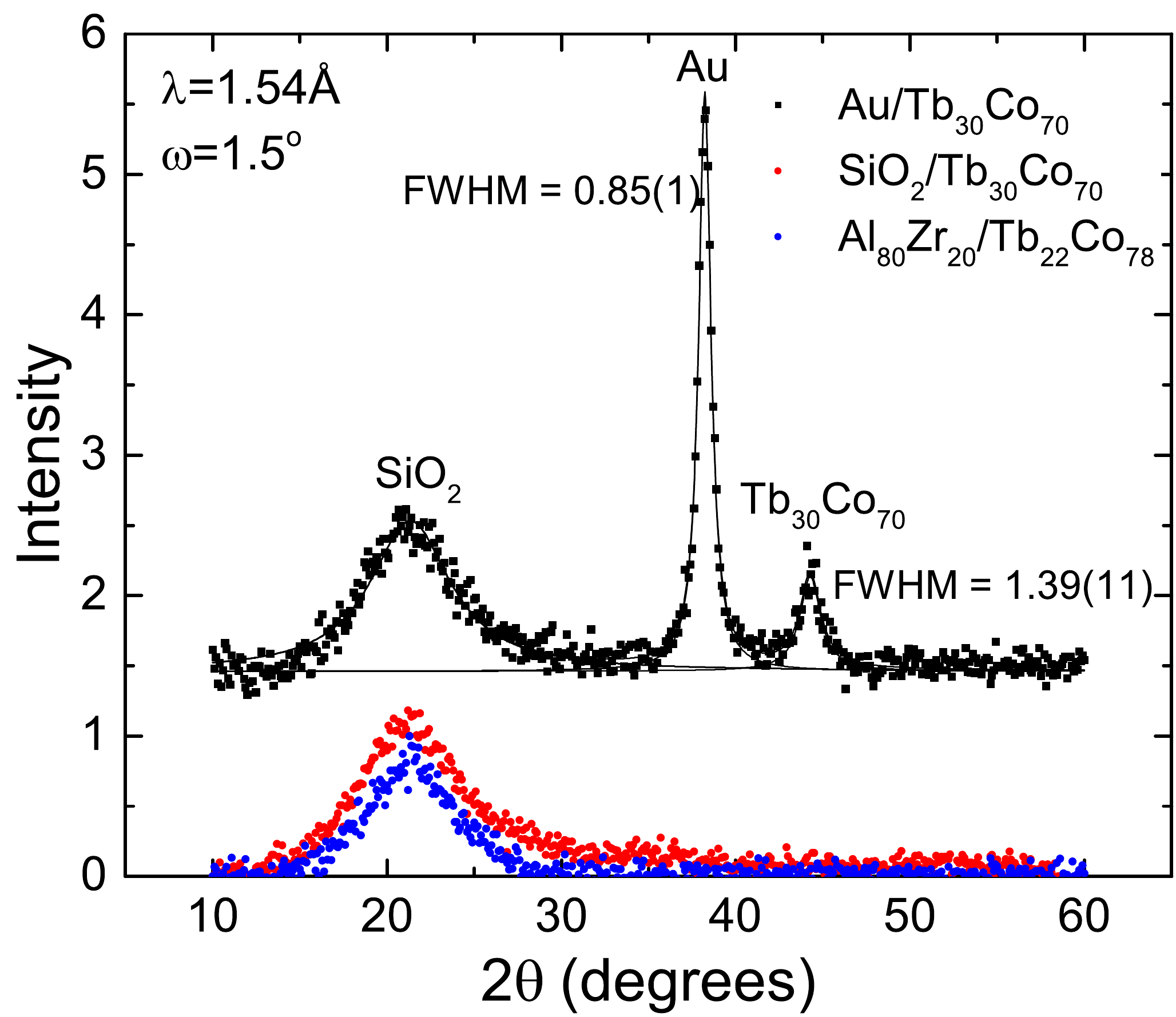}
    \caption{Diffractograms of a $\mathrm{Tb_{30}Co_{70}}$ sample deposited onto a fused silica substrate (red points) and onto a Au buffer layer (black points), both capped with $\mathrm{Al_{2}O_{3}}$, as well as of a sample deposited onto and capped with $\mathrm{Al_{80}Zr_{20}}$ (blue points).  Diffractograms were deconvoluted into Lorentzians, given by the solid black lines, and the intensity was normalized to the $\mathrm{SiO_{2}}$ peak. The data of $\mathrm{Au/Tb_{30}Co_{70}}$ sample (black measurement points) have been shifted upwards for clarity.}
    \label{fig:gixrd}
\end{figure}

The structure of the fabricated Tb$_{x}$Co$_{100-x}$ films prepared on fused silica substrates, without any buffer layer as well as with a Au or an $\mathrm{Al_{80}Zr_{20}}$, buffer layers, was investigated employing GIXRD measurements. The $\mathrm{Al_{80}Zr_{20}}$ buffer layers were used to ensure amorphicity of the Tb$_{x}$Co$_{100-x}$ films, as previously reported by \citet{Frisk_2015}. The Au buffer layers were chosen in this study in order to investigate influence on the structure and thus possibility of directly combining Tb$_{x}$Co$_{100-x}$ alloys with typical plasmonic materials such as Au, a prerequisite for the fabrication of future Tb$_{x}$Co$_{100-x}$-based magneto-plasmonic architectures. A typical GIXRD diffractogram for two $\mathrm{Tb_{30}Co_{70}}$ films, one prepared on a fused silica substrate and the other on a Au buffer layer, and for a  $\mathrm{Tb_{22}Co_{78}}$ film, prepared on an $\mathrm{Al_{80}Zr_{20}}$ buffer layer, are shown in Fig.\ \ref{fig:gixrd}. The films, prepared directly onto a fused silica substrate and on the $\mathrm{Al_{80}Zr_{20}}$ buffer layer, exhibit a single, low intensity broad peak at around $2\theta=22^o$, that can be associated with amorphous SiO$_2$ (red symbols), while three peaks can be observed for the film prepared onto a 20 nm thick Au buffer layer (black symbols). The first peak, at $2\theta=22^o$, similarly to the previous case, corresponds to the substrate; the second, narrow high-intensity peak around $2\theta=38^o$ arises due to the Bragg reflection from Au (111) planes, while the third, narrow low intensity peak corresponds to the $\mathrm{Tb_{30}Co_{70}}$ film.

In the work by \citet{Frisk_2015} it was shown that the amorphicity of Tb$_{x}$Co$_{100-x}$ films depends on the Tb content in the film and the crystallization onset is at around 80 at.$\%$ of Tb. This comparison shows that films deposited onto amorphous substrates, such as fused silica or $\mathrm{Al_{2}O_{3}}$, grow amorphous, while deposition onto the polycrystalline buffer layers such as Au, leads to a probable formation of nanocrystallites at the interfaces, similar to what has been reported by \citet{Liebig_2007cs}. Crystallite sizes, computed using Scherrer's equation, are 97.9 \AA~and 58.7 \AA~for Au and $\mathrm{Tb_{30}Co_{70}}$ layers, respectively. To further confirm amorphous structure of the Tb$_{x}$Co$_{100-x}$ films, we performed extended X-ray absorption fine structure (EXAFS) measurements of a Tb$_{x}$Co$_{100-x}$ film deposited on a fused silica substrate, which reveal no long range order of atoms in our films. A spectrum recorded with X-rays tuned for the Co edge is shown in Figure 2 of the Supplementary information.

\subsection{Magnetic and chemical characterization}

We investigated how the magnetic properties of Tb$_{x}$Co$_{100-x}$ films are affected by the different non-magnetic buffer layers, on which they are deposited, and imaged the remanent magnetization state.
 The effect of various buffer layers on the coercive field, $\mathrm{\mu_0H_c}$, of a 20 nm thick $\mathrm{Tb_{18}Co_{82}}$ film is illustrated in Supplementary information Fig. 3.  
 The $\mathrm{\mu_0H_c}$ is largest for samples prepared on an $\mathrm{Al_{80}Zr_{20}}$ buffer layer (364 mT) while the films deposited on a 20 nm thick Au buffer layer appear to have a reduced $\mathrm{\mu_0H_c}$ (170 mT). This can be related to the structure of the Tb$_{x}$Co$_{100-x}$ films: $\mathrm{Al_{80}Zr_{20}}$ is known to promote the amorphous growth of the films, while a polycrystalline Au buffer promotes nanocrystallite formation in TbCo films at the interface (see Fig. \ref{fig:gixrd}) \cite{Liebig_2007cs}. However, an increase in $\mathrm{\mu_0H_c}$ (to 280 mT) can be observed when a $\mathrm{Tb_{18}Co_{82}}$ film is deposited onto a hybrid buffer of $\mathrm{Au/Al_{2}O_{3}}$, so that $\mathrm{Tb_{18}Co_{82}}$ has an interface with amorphous $\mathrm{Al_{2}O_{3}}$. This demonstrates how the higher degree of amorphicity allows for increased coercivity in Tb$_{x}$Co$_{100-x}$ alloys. We would like to emphasize the impact of having a non-magnetic buffer layer which modifies atomic structure of the deposited magnetic layer and hence, its magnetic properties. This occurs without any interlayer exchange interactions, between the buffer layer and the magnetic active layer, which could be used for magnetic property modification, as in the study by \citet{tang_interfacial_2015}, where [Co/Ni] multilayers were used for tuning the $\mathrm{x_{comp}}$ of TbCo. 
 
\begin{figure}[!htbp]
  \centering
    \includegraphics[width=\columnwidth]{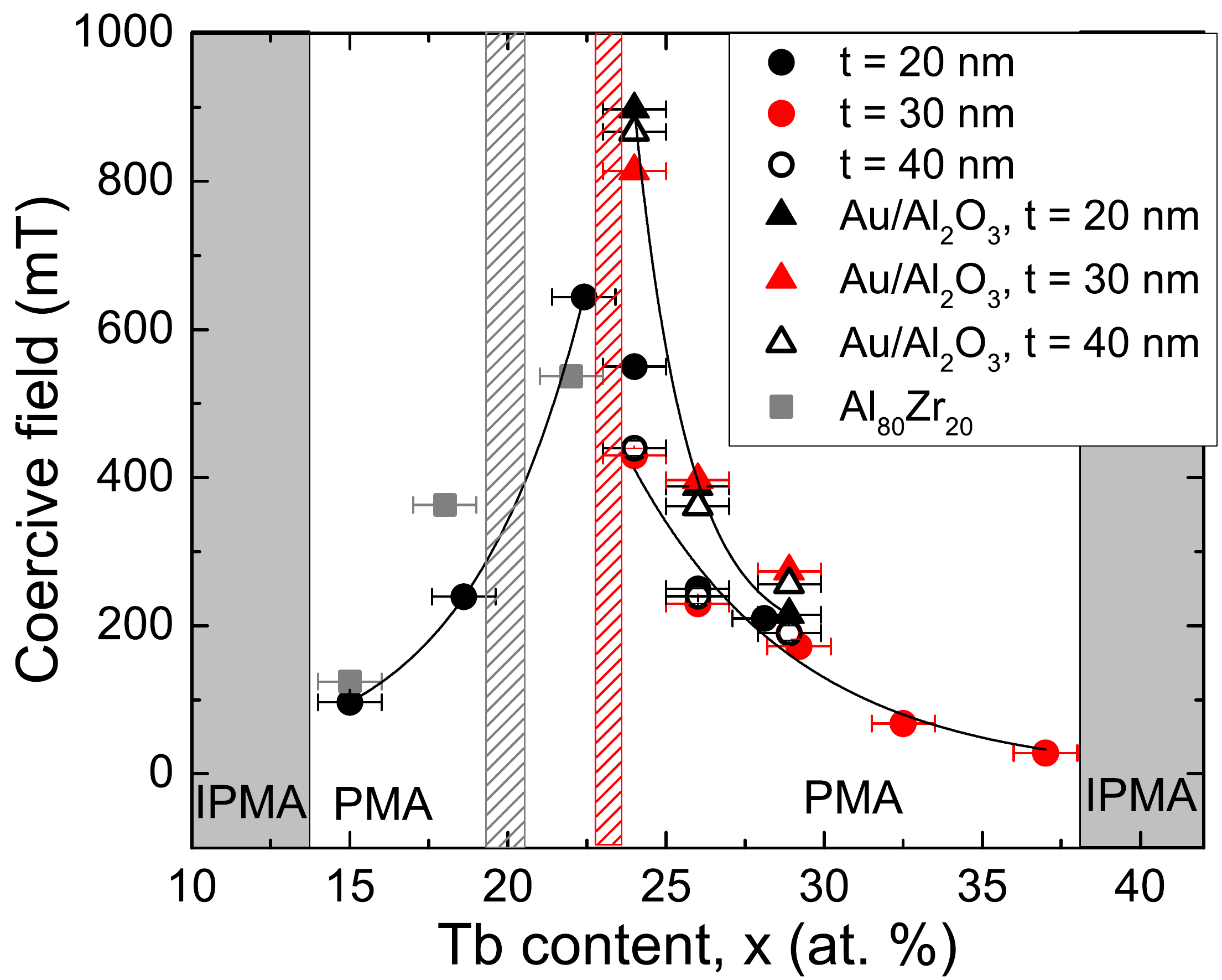}
    \caption{Coercive field dependence on Tb$_{x}$Co$_{100-x}$ composition for samples deposited on fused silica substrates (illustrated as circles), on fused silica/Au(20nm)/$\mathrm{Al_{2}O_{3}}$(3 nm) buffer structures (triangles) and on $\mathrm{Al_{80}Zr_{20}(3 nm)}$ buffer layers (squares). Black, red and empty symbols correspond to $t$ = 20, 30 and 40 nm thickness of Tb$_{x}$Co$_{100-x}$ layer, respectively. Correspondingly, IPMA and PMA denote regions, where the alloys exhibit an in-plane and perpendicular magnetic anisotropy. Red and grey hashed areas correspond to the $x\mathrm{_{comp}}$ for the sample series prepared directly on fused silica substrate and onto $\mathrm{Al_{80}Zr_{20}(3 nm)}$ buffer layers, respectively. Measurements were performed in polar MOKE configuration, with incident light of 530 nm wavelength.}
    \label{fig:Tb_content_Hc}
  \end{figure}

We further analyzed how the coercivity changes by varying composition and sample thickness. We varied the Tb content in the Tb$_{x}$Co$_{100-x}$ films of 20, 30 and 40 nm in thickness, deposited on untreated fused silica substrates and on a hybrid Au(20nm)/$\mathrm{Al_{2}O_{3}}$ buffer layers. The results are summarized in Fig. \ref{fig:Tb_content_Hc}. For films grown on the untreated silica substrates, the $\mathrm{\mu_0H_c}$ diverges at around 23 at.$\%$ Tb both from Co- and Tb- rich sides which is characteristic to ferrimagnetic alloys. This region, indicated by the red hashed area, corresponds to $x\mathrm{_{comp}}$ at room temperature. Similar values of $x\mathrm{_{comp}}$ at room temperature were reported in work by \citet{Alebrand_TbCo_magn_rev}. Grey regions at the lowest and highest Tb content represent ranges, where the films of the same thickness exhibits an in-plane magnetic anisotropy (IPMA) rather than out-of-plane, or perpendicular, magnetic anisotropy (PMA). Our investigations also show that, reduction of the thickness of TbCo films to below 10 nm, results in the emergence of in-plane magnetic anisotropy at the expense of the desired PMA (see Supplementary information). Similarly to observations of the buffer layer influence on $\mathrm{\mu_0H_c}$ (See Fig. 3 in Supplementary information), the Tb$_{x}$Co$_{100-x}$ films prepared on hybrid $\mathrm{Au(20nm)/Al_{2}O_{3}}$ buffer layers exhibit larger coercive fields than their counterparts prepared directly on fused silica substrates.

In relation to the seed layer effect investigation, we also measured samples with 15, 18 and 22 at.$\%$ Tb prepared onto $\mathrm{Al_{80}Zr_{20}}$ buffer layers. We observed that in contrast to films with 15 and 18 at.$\%$ Tb, the measured out-of-plane hysteresis loop of the sample with 22 at.$\%$ Tb appears to have changed the sign (see discussion in Supplementary Information). This indicates that the compensation point for films prepared onto $\mathrm{Al_{80}Zr_{20}}$ buffer layer is around 20 at.$\%$ Tb at room temperature, which is in agreement with the results of the previous study by \citet{Frisk_2015}. These results demonstrate the importance of an underlying buffer layer for the observed magnetic properties of the Tb$_{x}$Co$_{100-x}$ films. In addition to investigations of RT $x\mathrm{_{comp}}$, $T\mathrm{_{comp}}$ was determined by temperature-dependent polar MOKE measurements of the $\mathrm{\mu_0H_c}$ of Tb$_{x}$Co$_{100-x}$ films, prepared directly on the fused silica substrates. 
A Tb$_{18}$Co$_{82}$ film, deposited directly on a fused silica substrate, shows $T\mathrm{_{comp}}$ at around 170 K. For a film with 22 at.\ $\%$ Tb content, a slight increase of $\mathrm{\mu_0H_c}$ is observed close to RT, which indicates that $T\mathrm{_{comp}}$ is slightly higher than 300 K, while divergence of $\mathrm{\mu_0H_c}$ is not observed in the accessible temperature range ($100 - 300$ K) for Tb$_{24}$Co$_{76}$, which means that Tb$_{x}$Co$_{100-x}$ films with x $>$ 22 at.\ $\%$ Tb have a $\mathrm{T_{comp}}$ higher than 300 K.
These observations emphasize the influence of a buffer layer on the magnetic properties of the ferrimagnetic Tb$_{x}$Co$_{100-x}$ layers.

We imaged the magnetic domain structure of the 20 nm thick Tb$_{24}$Co$_{76}$ film, deposited on fused silica substrates, using Kerr microscopy and XPEEM. Kerr microscopy image for this film shows a broad size distribution of the magnetic domains (see the inset Fig. 7 in the Supplementary information). Fig.\ \ref{Domains_PEEM}(a) shows X-ray absorption (XA) spectra recorded by means of XPEEM at the Co $L\mathrm{_{2,3}}$ edges. The spectral shape corresponds to metallic cobalt without trace of oxidation \cite{Regan-XAS}. Fig.\ \ref{Domains_PEEM}(b) shows the corresponding XA spectra at the Tb $M\mathrm{_5}$ edge. Magnetic contrast maps recorded at the Co $L\mathrm{_{2,3}}$ and the Tb $M\mathrm{_5}$ edges are shown in Fig.\ \ref{Domains_PEEM}(c) and (d), respectively, with domain sizes in the micrometer range. The contrast reversal between the domains visualized at the Co edge and the Tb edge directly confirms the ferrimagnetic coupling within the alloys. 

\begin{figure}[ht!]
    \centering
    \includegraphics[width=1\columnwidth]{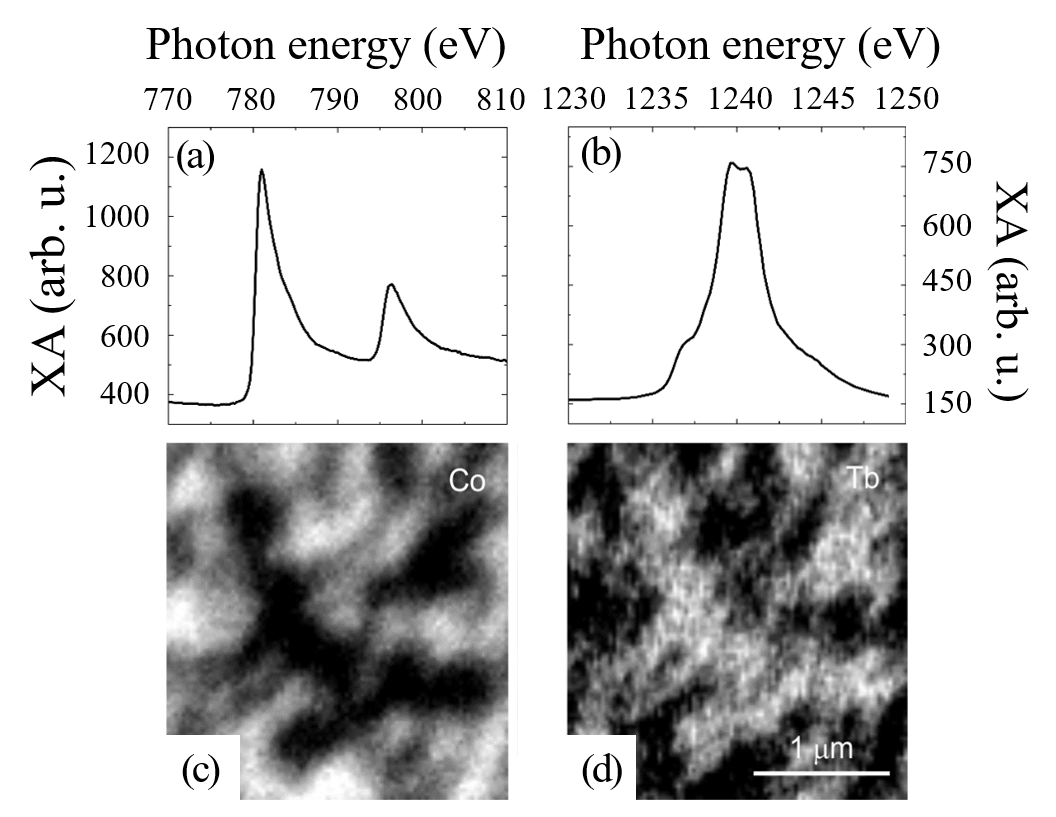}
    \caption{X-ray absorption spectra measured at the (a) Co $L\mathrm{_{2,3}}$ and (b) Tb $M\mathrm{_5}$ absorption edges and magnetic contrast maps of a remanent magnetization state of a 20 nm thick $\mathrm{Tb_{24}{Co}_{76}}$ film obtained by combining XPEEM with XMCD at the (c) Co $L\mathrm{_3}$ and the  (d) Tb $M\mathrm{_5}$ edges.}
    \label{Domains_PEEM}
  \end{figure}

\subsection{Optical characterization}

\begin{figure}
   \centering
    \includegraphics[width=\linewidth]{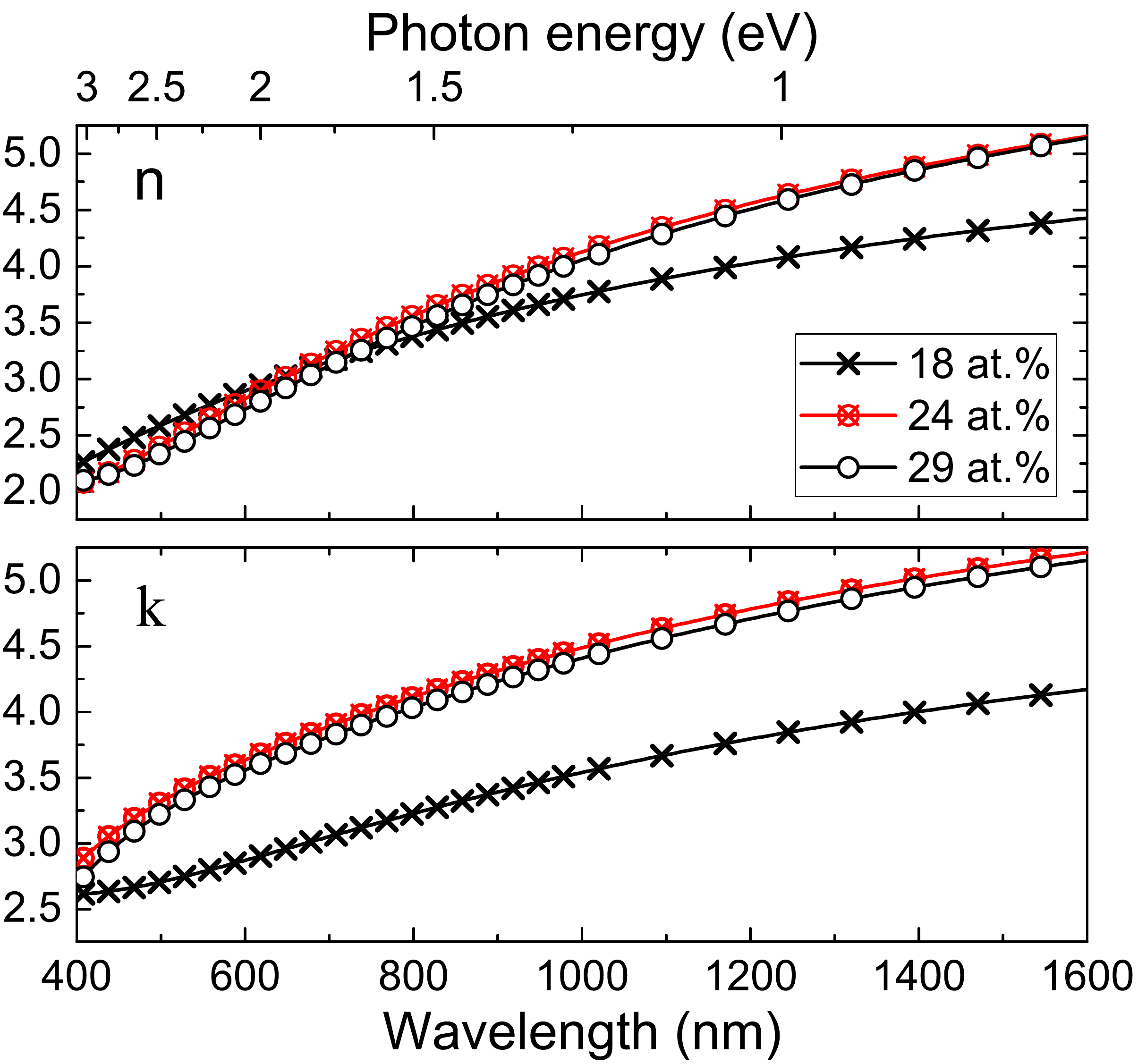}
  \caption{Dispersion of the refractive index, {\textit{n}}, and the extinction coefficient, {\textit{k}}, for various Tb contents in a film, obtained from fitting ellipsometry data in the visible and near-infrared range. Layer structure and thickness were taken into account for ellipsometry data fits.}
    \label{fig:n_k}
\end{figure}

The wavelength dependence of the refractive index, \textit{n}, and extinction coefficient, \textit{k}, extracted from the ellipsometric data, is summarized in Fig. \ref{fig:n_k} for three alloy films with different Tb content within the 18-30 at.\% range. For all samples, a monotonic increase with increasing wavelength is observed for both quantities. Similar qualitative behavior has also been recently reported for even higher concentrations of Tb  \cite{iemoto_interference_2018}. It appears that there is no significant dependence on the Tb content for samples with a Tb content above the $x\mathrm{_{comp}}$. However, for a sample with Tb content below compensation point, the {\textit{n}} and {\textit{k}} dispersion curves lie below the curves of samples with Tb content $x>x\mathrm{_{comp}}$. Once again, the optical constants determined here have been targeted having in mind the potential of Tb$_{x}$Co$_{100-x}$ films in magneto-plasmonic architectures (see e.g. the work by \citet{liu_nanoscale_2015},  \citet{Rich_Truncated_nanocones} and \citet{Feng_AdvOpt_2020}). For the proper electromagnetic design and simulations of these, knowledge of these optical constants is of paramount importance.

\subsection{Magneto-optical Imaging: All-optical Helicity Dependent Switching in the Tb$_{x}$Co$_{100-x}$ layer}

\begin{figure}[htpb!]
    \centering
    \includegraphics[width=\columnwidth]{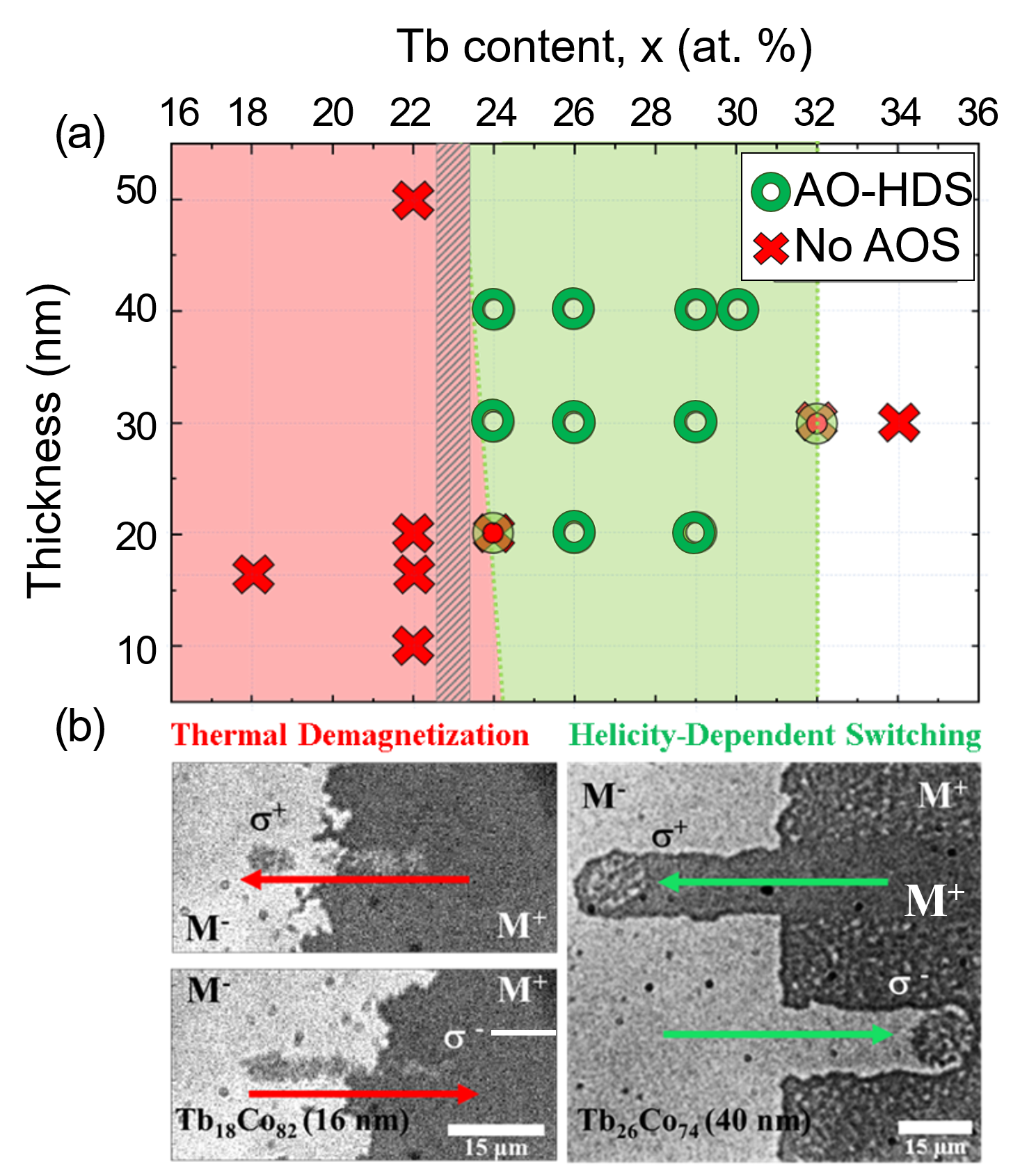}
    \caption{(a) Diagram showing the switching behavior as a function of sample thickness and composition. The red and green areas correspond to pure thermal demagnetization and multi-shot helicity dependent switching, respectively. In the white region, out-of plane domain contrast can no longer be resolved with our microscope. The grey hashed area corresponds to the compensation point $x_{\mathrm{comp}}$. (b) Helicity-independent thermal demagnetization as seen in the red region (left) and HD-AOS as seen in the green-shaded region of the diagram (right).}
    \label{fig:NoBuffer_PhaseDiag_AOS_RU}
\end{figure}

The conditions for observation of AOS were investigated in the Tb$_{x}$Co$_{100-x}$ samples with varying $x$, deposited directly on fused silica substrates and capped with $\mathrm{Al_{2}O_{3}}$ 3 nm layer \textit{in situ}. The results of multi-shot and sweeping-beam experiments confirm the cumulative mechanism of HD-AOS in TbCo alloys, which is also confirmed by the theoretical calculations, discussed towards the end of this section. Single-shot helicity-independent switching of magnetization was not observed in any of the samples studied. However, a range of samples do exhibit a multi-shot HD-AOS under suitable laser pulse parameters, such as fluence and sweeping speed. The range of concentration and thickness, where HD-AOS was observed (for these samples), is plotted in the diagram of Fig.\  \ref{fig:NoBuffer_PhaseDiag_AOS_RU}(a). Below 24 at.$\mathrm{\%}$ Tb content, thermal demagnetization is always observed. This is shown in the left panel of Fig.\ \ref{fig:NoBuffer_PhaseDiag_AOS_RU}(b), where both of the incident helicities, $\mathrm{\sigma^{+}}$ and $\mathrm{\sigma^{-}}$, induce a contrast change in two oppositely-oriented magnetic domains to an intermediate gray, indicating that both helicities result in demagnetization.

\begin{figure*}[ht!]
    \centering
    \includegraphics[width=\textwidth]{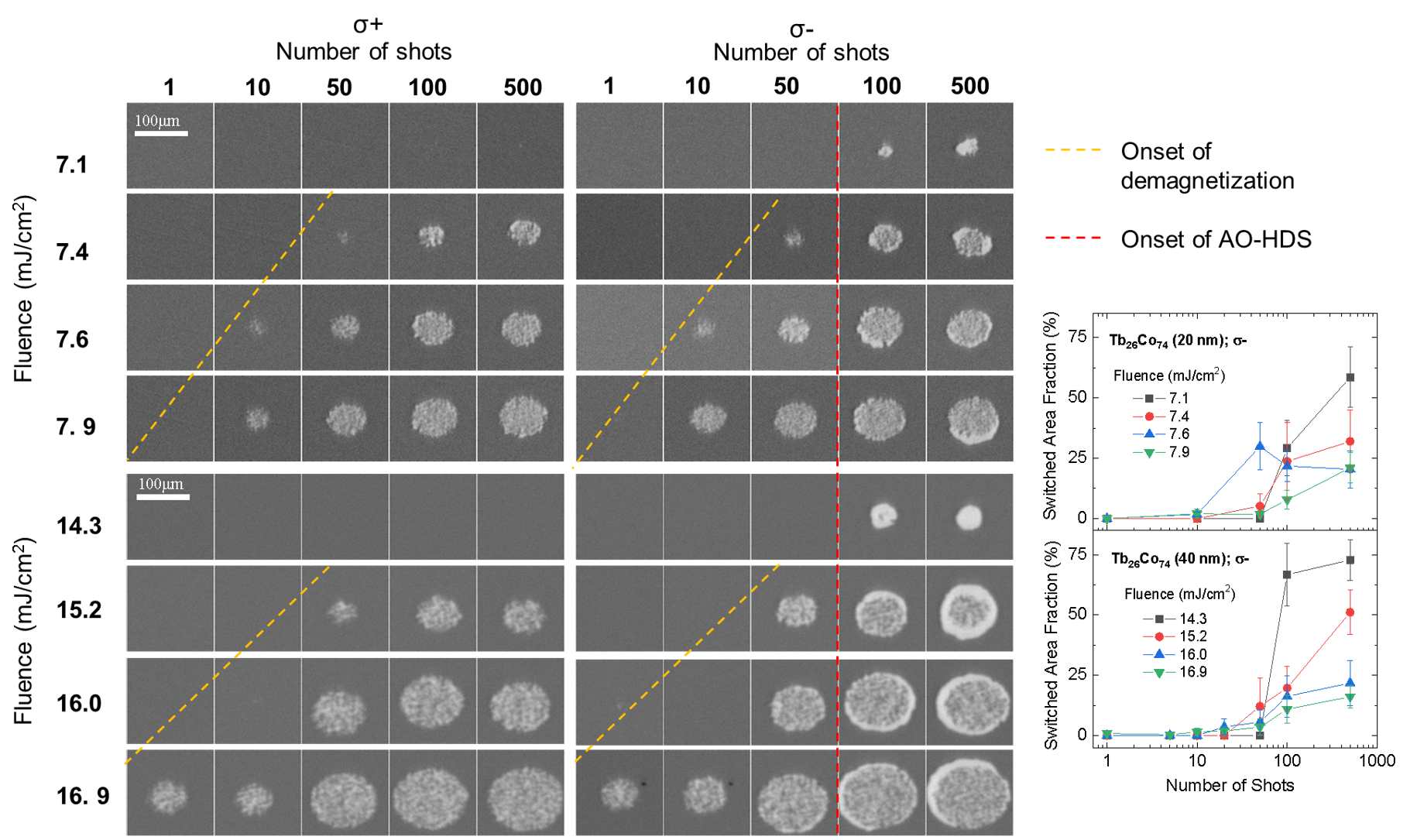}
    \caption{Observation of HD-AOS dependence on fluence and number of shots in multi-shot experiments on $\mathrm{Tb_{26}Co_{74}}$ films of thickness (a) 20 and (b) 40 nm in response to a train of pulses incident at the same spot on the sample. A ring of switched magnetization appears outside the central demagnetized region for the helicity $\mathrm{\sigma^-}$ for bursts containing 100 or more pulses. The inset shows the efficiency of AOS of magnetization as a ratio of the switched area to the total laser-affected area as a function of number of shots for different fluence for $\mathrm{\sigma^-}$ helicity.}
    \label{fig:Multi_pulse_RU}
\end{figure*}

Samples with Tb content ranging from 24 to 32 at.\% show HD-AOS, as seen in the right panel of Fig.\ \ref{fig:NoBuffer_PhaseDiag_AOS_RU}(b). For these samples, only one helicity can induce switching for a domain of a given orientation. The opposite helicity leaves the domain unaffected, showing a clear helicity-dependence of domain writing. Further, for samples with \textit{x}~$>$~32 at.$\%$, domain contrast cannot be observed with the polar Faraday imaging geometry, due to the shifting of easy out-of-plane magnetization axis towards the in-plane direction. Hence, it is unresolved whether HD-AOS occurs in this composition range or not. The laser pulse width used to test for AOS varies from 100 fs to 240 fs, while the laser spot diameter varies from 50 to 80 $\mu \mathrm{m}$. No qualitative change in the switching behavior was observed for these samples using these laser parameters. It can also be seen from Fig.\ \ref{fig:NoBuffer_PhaseDiag_AOS_RU}(a) that HD-AOS is not as strongly dependent on sample thickness as it is on the sample composition. Comparing these observations with Fig.\ \ref{fig:Tb_content_Hc}, which shows that at room temperature $x_{\mathrm{comp}}\sim 23$ at.\%, we can conclude that HD-AOS is observed only for samples with $x>x_{\mathrm{comp}}$, i.e.\ for the samples with $T\mathrm{_{comp}>}$ 295 K. This finding agrees with other studies, e.g.\ by \citet{Alebrand_TbCo_magn_rev, Mangin}, that have investigated switching in TbCo alloys as a function of Tb content, and found HD-AOS for samples for which 300 K $<\mathrm{T_{comp}<T_{Curie}}$, where $\mathrm{T_{Curie}}$ is the Curie temperature. 
Thus the film composition, with respect to the $x\mathrm{_{comp}}$, clearly has a strong impact on the possibility of achieving HD-AOS.

Some representative results of multi-pulse switching in $\mathrm{Tb_{26}Co_{74}}$ films of two different thicknesses are summarized in Fig.\ \ref{fig:Multi_pulse_RU}. A specific threshold fluence, below which no contrast changes, i.e.\ no switching or demagnetization, can be observed for all investigated samples (Dependence of threshold fluence for samples with different Tb-content is shown in Supplementary information, Fig. 9.). Therefore, in Fig.\ \ref{fig:Multi_pulse_RU} we show only the results of measurements performed with a fluence slightly above the threshold fluence. A demagnetized region, clearly resolvable only at a fluence much higher than the threshold fluence, appears with a single laser shot for 40 nm thick film. As the incident fluence is increased, this demagnetized region grows larger but no HD-AOS is seen for a single shot of any fluence, relating to recent observations made by \citet{Quessab_PhysRevB_2019}. This behavior is also seen for trains of 10 and 50 pulses incident on the sample. Due to the larger number of shots, the demagnetized region grows larger and becomes resolvable for even smaller fluences. The yellow dashed line illustrates how the demagnetization region becomes resolvable at lower fluences for a larger number of shots. The combination of the fluence and number of shots determines the heat deposited to the sample, which in turn determines the size of the demagnetized region. However, no HD-AOS, only demagnetization, is observed below 100 shots, irrespective of the incident fluence, for either of the samples studied here.

Close to the threshold fluence, pure helicity-dependent switching is observed for all the samples i.e.\ for one helicity, a complete reversal of magnetization is observed without any central demagnetized region, whereas no contrast change is induced by the opposite helicity. These results further elucidate the effect of film thickness on the switching behavior. Though the film thickness does not play a crucial role in determining the presence of HD-AOS for the samples investigated in this study, it definitely affects the `quality of switching'. While both 20 and 40 nm thick $\mathrm{Tb_{26}Co_{74}}$ samples show an outer rim of HD-AOS outside the demagnetized region, the rim is much more prominent, thicker and homogeneously shaped for the 40 nm thick films. This further supports the observation of better quality of switching for thicker films, possibly due to better heat dissipation upon illumination with a laser pulse, observed in beam sweeping experiments 
\cite{banerjee2019single}.

\begin{figure}[ht!]
    \centering
    \includegraphics[width=\columnwidth]{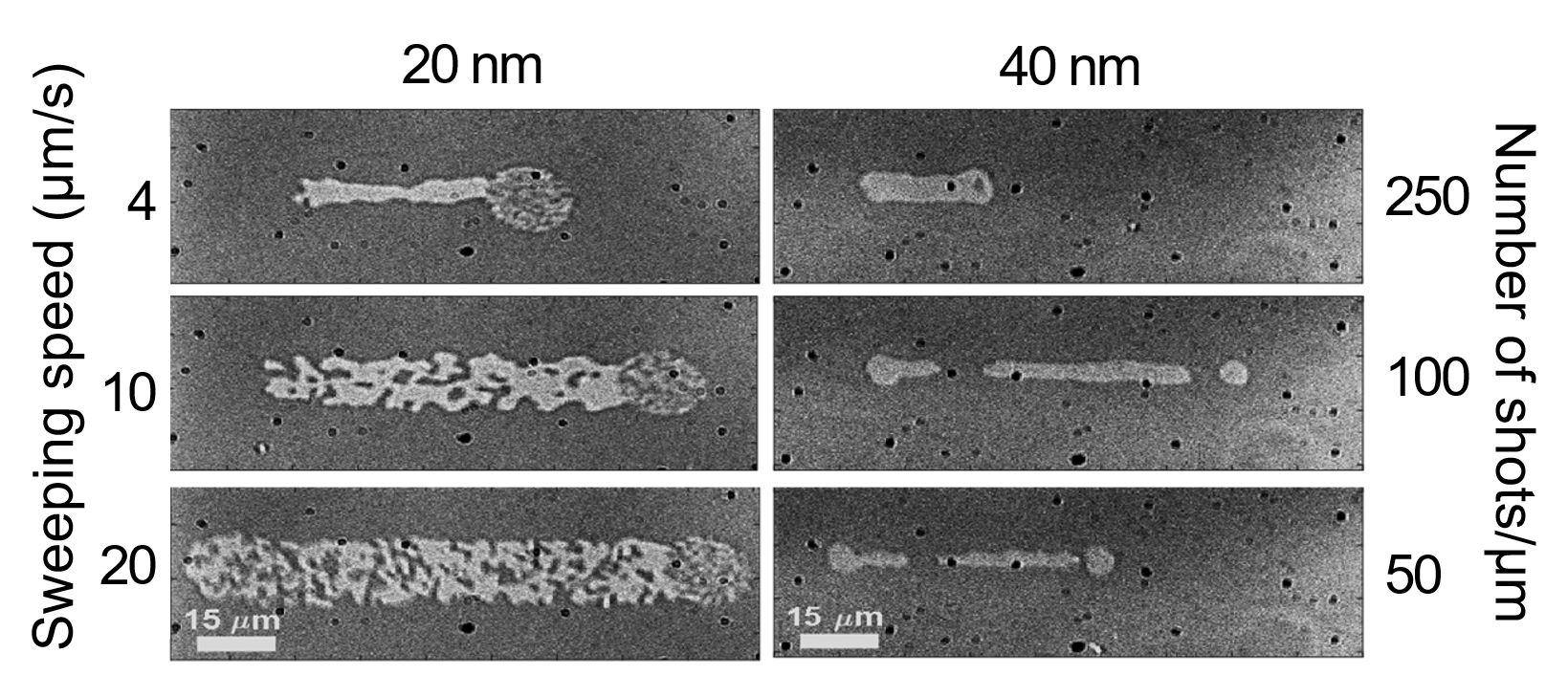}
    \caption{Sweeping speed dependence of HD-AOS in beam-sweeping experiments on $\mathrm{Tb}_{26}\mathrm{Co}_{74}$ films of (left) 20 and (right) 40 nm thickness.}
    \label{fig:Fluence_Speed_Dep_RU}
\end{figure}

Results of the beam-sweeping experiments on the  20 nm and 40 nm thick $\mathrm{Tb}_{26}\mathrm{Co}_{74}$ films are shown in Fig.\ \ref{fig:Fluence_Speed_Dep_RU}, which illustrates the dependence of the observed HD-AOS on the beam sweeping speed. It shows how, on increasing the sweeping speed, the switched area transitions from a homogeneously switched region to discontinuously switched areas. This supports the multi-shot switching mechanism further. A higher speed correlates to fewer shots per unit length (since the repetition rate of the shots is constant but the beam moves over the sample faster) and thus the discontinuity in switching can be explained due to fewer number of shots and thus smaller fluence received by a unit area. It can be seen from the results obtained for two film thicknesses that the fraction of the laser-swept area that undergoes homogeneous switching is larger for 20 nm thickness as compared to 40 nm, for the same film composition.

We conclude that thickness does not play a crucial role in determining the switching behavior for the 20 to 40 nm thick films. \citet{hadri_domain_size_criterion_PhysRevB.94.064419} studied the role of thickness in HD-AOS in TbCo films for a thickness range of 1.5 nm to 20 nm. In this range, they found that the switching behavior was strongly influenced by the film thickness. This was attributed to the sharp variation of domain sizes as a function of film thickness in this critical regime. Their calculations show that the variation of domain size with film thickness, is not as drastic in the range of $20-40$ nm. The size of a magnetic domain was considered as one of the criteria for the observation of HD-AOS in both ferromagnetic Co/Pt and Co/Ni films, as well as in ferrimagnetic TbCo alloys of certain thicknesses \cite{hadri_domain_size_criterion_PhysRevB.94.064419}. More specifically, it was discussed that HD-AOS can be achieved in films of thickness $1.5 - 6$ nm with 8 to 15 at.$\%$ of Tb, or films of thickness $3.5 - 20$ nm with $15 - 30.5$ at.$\%$ of Tb, and defined a criterion for the observation of AOS to be the equilibrium domain size after laser-induced heating which becomes larger than the laser spot size. Our findings show that HD-AOS can be achieved using suitable laser pulses in thicker TbCo films, which have a composition above the compensation point, and have randomly shaped domains with a broad size distribution from around 40 $\mu$m for 20 nm thick film with 24 at.\% Tb to 10 $\mu$m and smaller domains for 30 nm thick films with $24 - 28$ at.\% Tb. The domain sizes are either comparable or smaller, depending on the composition (and thickness), with our laser spot size of $50 - 80$ $\mu m$. We observed that even the sample with a composition below the $x_{\mathrm{comp}}$, i.e. 20 nm thick $\mathrm{Tb}_{26}\mathrm{Co}_{74}$, has an average domain size within the range of the domain sizes of the samples with compositions larger than the compensation point. However, it exhibits only helicity-independent thermal demagnetization and not a HD-AOS. Therefore, in our study, the domain size with respect to the laser spot size does not appear to be a determining parameter for the observation of AOS. Instead, we explain the observation of helicity-dependent AOS in our films with the help of theoretical calculations of the IFE. 
\begin{figure}[t!]
    \centering
    \includegraphics[width=\columnwidth]{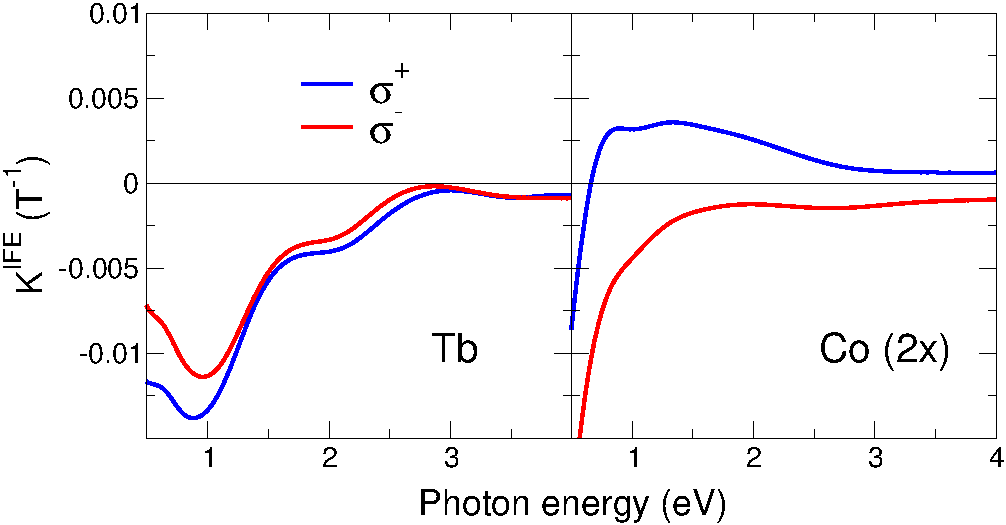}
    \caption{\textit{Ab initio} computed helicity- and frequency-dependent IFE constants $K^{\rm IFE}$ of Tb and Co atoms in TbCo$_2$. Note that the values for Co are multiplied by a factor of two.}
    \label{fig:IFE_theory}
\end{figure}

\subsection{{\it Ab initio} calculations of the inverse Faraday effect}

The \textit{ab initio} computed frequency-dependent IFE constants for the Tb and Co atoms are shown in Fig.\ \ref{fig:IFE_theory}. Note that in the here-studied excitation geometry the wave vector of the circularly polarized light is parallel to the TbCo magnetization, to simulate a TbCo film with PMA and illuminated  at normal incidence. We adopt as model system the ordered ferrimagnetic Tb$_{33}$Co$_{67}$ compound, whose composition is on the edge of the AOS region shown in Fig.\ \ref{fig:NoBuffer_PhaseDiag_AOS_RU}(a). In the \textit{ab initio} calculations the Tb atoms have a 9.1 $\mu_{\rm B}$ magnetic moment {($\mu_s + \mu_l$)} and the Co atoms an antiparallel $1.3$ $\mu_{\rm B}$ moment.
Further, we mention that to obtain the laser-induced moment per atom the results given in Fig.\ \ref{fig:IFE_theory} are to be multiplied by the unit cell volume and by $I/c$ (for details, see \cite{Berritta2016}).  

As can be directly recognized, the induced magnetic moment on the Tb atom displays practically no helicity dependence for a photon energy of 1.55 eV. Consequently, on the Tb atom there will be a laser-induced magnetic moment but it will be antiparallel to the Tb atomic moment irrespective of the light's helicity. This in turn implies that helicity-dependent switching of the large Tb moment is improbable. The situation is however different for the Co atom. At around 1.55 eV photon energy the induced magnetic moment is strongly helicity dependent and, importantly, changes sign when the helicity is reverted, which is favorable for helicity-dependent AOS. The helicity-dependent moments induced on the Co atoms can initiate a magnetization reversal, but, as its influence is small (about $\pm 0.05$ to $\pm 0.10$ $\mu_{\rm B}$ for the here-used experimental fluences), it requires the accumulation of many laser shots in a stochastic process to bring about the magnetization switching. To study the full process of both laser-induced demagnetization and switching it would be required to perform magnetization dynamics simulations (see, e.g., \cite{Wienholdt2013,John2017}).  

Our first-principles calculations hence show that in particular the helicity-dependent laser-induced moments on the Co atoms are responsible for causing HD-AOS in TbCo, refining recently reported switching mechanisms \cite{Quessab_PhysRevB_2019}. A complete picture of helicity-independent AOS and HD-AOS in RE-TM alloys however still needs to be achieved. Thus far, helicity-independent, single-shot switching was unambiguously observed for GdFeCo (with $\sim$8\% Co) \cite{Stanciu_AO_circularly_polarized_light}. Recently, single pulse switching was observed, too, for a Gd/Co bilayer \cite{Lalieu2017}. An ingredient for this switching behavior could be the relatively slow response of the Gd $4f$ magnetic moment upon laser excitation, since the deep-lying $4f$ electrons are not excited by the pump laser, in contrast to the instantaneous excitation of spin-polarized itinerant $3d$ electrons. Recent magnetic dichroism in photoemission experiments measured a slow demagnetization behavior of the Gd $4f$ moment in pure Gd ($\sim$14 ps) but a subpicosecond demagnetization of the $4f$ moment in Tb \cite{Frietsch2015,Frietsch2020}. This gives an indication that the different intrinsic response times of the $4f$ moments in Gd and Tb could play a role for the AOS. However, X-ray magnetic circular  dichroism (XMCD) experiments performed on Gd and Tb observed both a fast and a slow $4f$ demagnetization in Gd and in Tb metal \cite{Wietstruk2011}. L{\'o}pez-Flores \textit{et al.}\ \cite{Lopez2013} used XMCD to measure the element-dependent demagnetization rates in GdCo and TbCo alloys and found subpicosecond demagnetization rates for both Gd and Tb. These different probing techniques thus provide rather disparate values for the $4f$ demagnetization rates and the origin of this discrepancy will need to be clarified in order to resolve fully why these RE-TM alloys exhibit a distinct AOS behavior. To complicate the matter further, very recently a first observation of single shot AOS was reported for Tb/Co multilayer stacks \cite{aviles-felix_single-shot_2020}. Such behavior is however not found in our investigation of amorphous TbCo alloys and was neither found in previous studies \cite{ElHadri2016,Hadri-review}.

\section{Discussion and Conclusions}

We determined the composition and thickness ranges where the Tb${_x}$Co$_{100-x}$ films exhibit perpendicular magnetic anisotropy, reporting that the magnetization compensation point at room temperature is around 23 at.$\%$ Tb for the samples deposited onto a fused silica or onto $\mathrm{Al_2{O}_3}$ buffer layers, while the samples deposited onto $\mathrm{Al_{80}{Zr}_{20}}$ have a compensation point around 20 at.$\%$ Tb. When deposited onto amorphous layers such as fused silica and alumina, Tb${_x}$Co$_{100-x}$ films are amorphous, 
whereas there are signatures of nanoscrystallite formation in the films deposited onto polycrystalline Au buffer layers. 

We further explored our parameter space for an optimal region for observation of HD-AOS. We find that Tb${_x}$Co$_{100-x}$ alloys which have $x>x\mathrm{_{comp}}$ at room temperature, deposited directly onto fused silica substrates exhibit HD-AOS. For these amorphous films, the sample composition is the most important parameter determining the switching behavior, and HD-AOS was observed for the films with compositions in the range of $24 - 30$ at.$\%$ of Tb content, while the remaining samples with lower Tb content show thermal demagnetization. The samples with $x$~$>$~32 at.$\%$, exhibit in-plane magnetic anisotropy and therefore no AOS was observed due to our experimental setup, configured for the out-of-plane magnetization switching measurements. Helicity dependence shows up in the sample's response to a train of ultrashort laser pulses only after $50 - 100$ shots, with a weak dependence on the applied laser fluence. The film thickness furthermore has a limited influence on HD-AOS, only in terms of raising the threshold fluence, above which either HD-AOS or demagnetization is observed, and by improving the `quality' of the observed HD-AOS. Lastly, our first-principles calculations explain how the AOS can occur in TbCo. We showed that circularly polarized laser light induces a helicity dependent magnetic moment on the Co atoms that in particular can trigger the HD-AOS in a cumulative, multi-shot process. 

This study shows that ferrimagnetic Tb${_x}$Co$_{100-x}$ films in a suitable range of compositions can exhibit HD-AOS under a range of controlled experimental parameters, such as a laser fluence and sweeping speed. We also show how to control the properties like the coercive field, switching behavior and room temperature compensation point, while maintaining an out-of-plane magnetic anisotropy by growing these layers onto a variety of buffer layers. Our observations underline the potential of the combination of TbCo with suitable plasmonic materials in order to fabricate hybrid magneto-plasmonic architectures \cite{Rich_Truncated_nanocones}, possibly enhancing the magneto-optical activity and ultrafast AOS, as was recently shown by \citet{liu_nanoscale_2015} using patterned Au antennas onto a TbFeCo film, or \citet{Feng_AdvOpt_2020} for Co/Pt with Au. Such architectures are potential candidates for the future magnetic memory storage devices based on opto-magnetic effects \cite{Nanoscale_Magnetophotonics_JAP2020}.

The data that support this study are available via the Zenodo repository \cite{zenodo}.

\begin{acknowledgements}
The authors acknowledge support from the Knut and Alice Wallenberg Foundation project ``{\it Harnessing light and spins through plasmons at the nanoscale}'' (No. 2015.0060), the Swedish Research Council (Project No. 2019-03581) and the Swedish Foundation for International Cooperation in Research and Higher Education (Project No. KO2016-6889). This work is also part of a project which has received funding from the European Union's Horizon 2020 research and innovation program under grant agreement No. 737093, ``{\textsc{femtoterabyte}}",
and has been supported by the Swedish National Infrastructure for Computing (SNIC).

The operation of the Tandem Accelerator Laboratory has been supported by infrastructural grants from the Swedish Foundation for Strategic Research (SSF-RIF14-0053) and the Swedish Research Council (contracts \#~821-2012-5144 and \#~2017-00646-9). 

K.M.\ and A.K.\ (Andrei Kirilyuk) would like to acknowledge Guido Bonfiglio, Dr.\ Kihiro Yamada for assistance with experimental setups for imaging and temperature-dependent hysteresis measurements, and Dr.\ Sergey Semin and Chris Berkhout for technical support. M.V.M.\ would like to acknowledge Dr.\ Daniel Primetzhofer for discussions in the ion beam analysis section. A.C., G.A.\ and V.K.\ would like to thank Dr.\ Maarten Nachtegaal and SuperXAS - X10DA beamline at PSI for providing beamtime for EXAFS measurements, and Sebastian George for help in the beamtime and data analysis, as well as for providing initial samples for AOS testing. Dr.\ Gunnar K.\ P\'alsson is acknowledged for help with computational aspects of the domain size evaluation and GIXRD.
\end{acknowledgements}


%

\cleardoublepage
\section*{Supplementary Information}

\section{Sample preparation and characterization}
\subsection{Co-sputtering}
Films were prepared by co-sputtering Tb and Co in ultrahigh vacuum chamber with a base pressure of $10^{-10}$ -$10^{-9}$ Torr, while the Ar$^+$ sputtering gas pressure was 2-3 mTorr. The Tb:Co ratio was varied by keeping a constant power on Tb target and varying power on Co target. The buffer and capping layers of $\mathrm{Al_{80}{Zr}_{20}}$ and $\mathrm{Al_2O_3}$ were prepared \textit{in situ}.

\subsection{Elemental characterization}
For the RBS and PIXE measurements, the films were loaded on a wheel-sample holder mounted on a computer-controlled goniometer, which allows simultaneous data acquisition and sample positioning. The incident/exit angles of the sample were automatically "randomized" during the measurements by small steps ($\pm 2\degree$) around an equilibrium incident angular position ($\alpha = 5\degree$) in order to reduce residual channeling effects. For the RBS, a Passivated Implanted Planar Silicon (PIPS) detector with  energy resolution of FWHM $\approx$ 13 keV (for the whole detection chain) was placed at a scattering angle of $\theta$ = 170$\degree$. The RBS measurements were carried out in a low current regime, aiming to avoid problems with pile-up effects ($<$1$\%$ for all the measurements). The experimental RBS spectra were analyzed using the latest version of the SIMNRA code~\cite{Mayer_SIMNRA7-1}, and the stopping power data used as input in the fits was selected from the last version of SRIM~\cite{Ziegler-2} code. Simultaneously with RBS measurements, PIXE spectra were recorded for each sample, using a silicon drift detector (SDD) placed at $\theta$ = 135$\degree$ with a 79.5 $\mu$m Mylar absorber in front of the Be-window (resolution: FWHM $\approx$ 143 eV for Fe-K$_\alpha$ characteristic energy). The PIXE spectra were fitted using the GUPIX code~\cite{Campbell-3}. Both RBS and PIXE spectra were analyzed following an iterative and self-consistent approach (further details of the setup and data analysis routine can be found in~\cite{MORO2019137416}).

\subsection{Structural characterization}


\begin{figure}[h!]
	\centering
	\includegraphics[width=0.9\columnwidth]{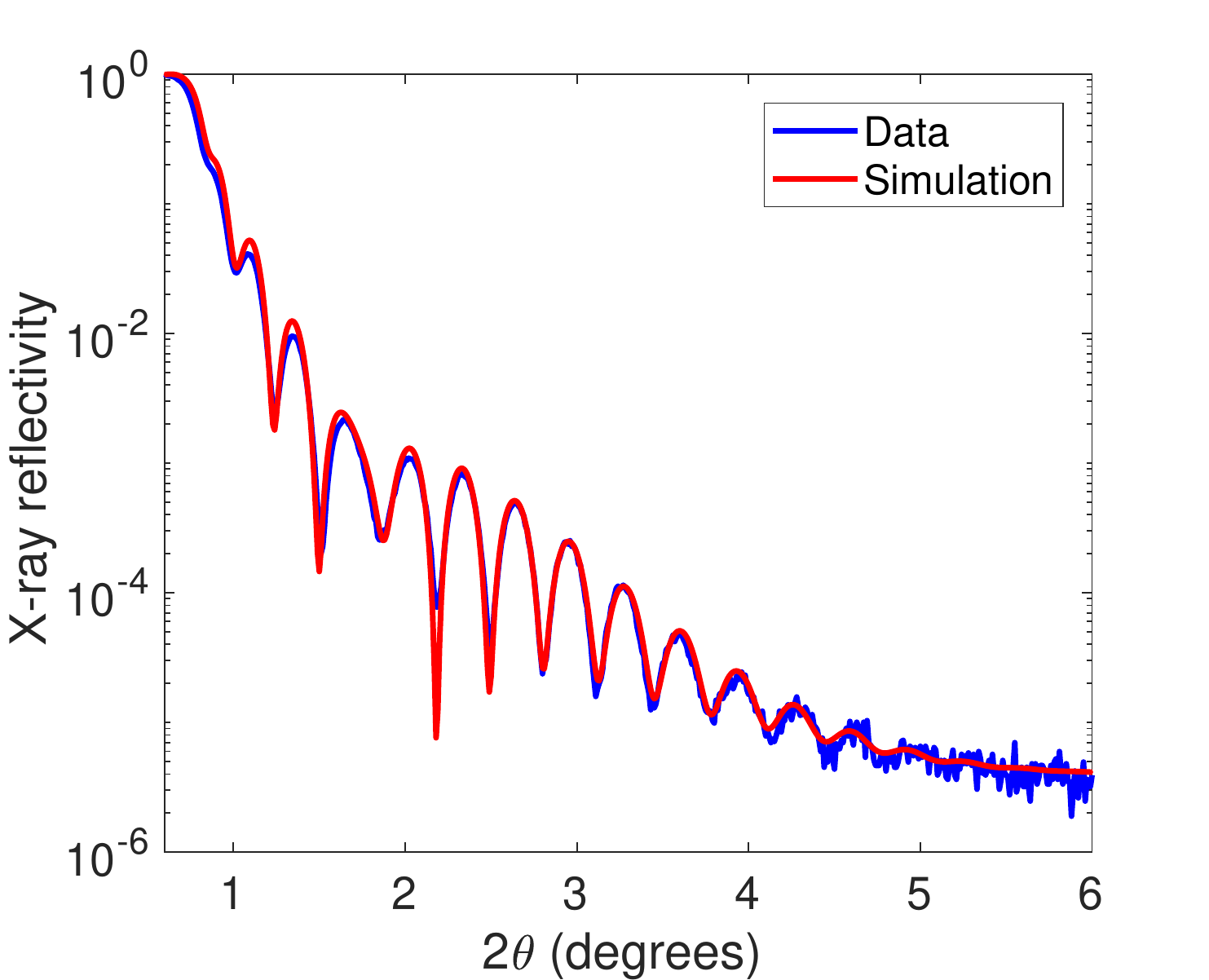}
	\caption{XRR of $\mathrm{Tb_{24}Co_{76}}$ film, deposited onto a fused silica substrate and capped with $\mathrm{Al_2O_3}$, measured with x-rays of 1.5418 \AA~ in wavelength. From fitting the x-ray data in  GenX we extracted the following thickness: 260.7(32) \AA and 31.2(66) \AA for $\mathrm{Tb_{24}Co_{76}}$ and $\mathrm{Al_2O_3}$ layers, respectively. The fitted TbCo layer roughness is 5.2 \AA.}
	\label{fig:XRR}
\end{figure}

The thickness and roughness of individual layers comprising our structures, were determined using x-ray reflectivity (XRR) and fitting the obtained data to a \textit{GenX} model \cite{GenX} corresponding to the actual layer structure. An example of measured data and corresponding fit is shown in Fig. \ref{fig:XRR}, for a sample of the following structure: fused silica/$\mathrm{Tb_{24}Co_{76}/Al_2O_3}$. A summary with selected TbCo samples, prepared on a range of buffer layers, is shown in the Table \ref{tab:1}. Grazing incidence x-ray diffraction (GIXRD) measurements were performed to determine the layered structure of the films. The x-ray beam was incident at grazing angles in the range of $\omega$ = 0.5-1.5$\degree$, while the detector scanned the range $2\theta$ = 10-60$\degree$.

\begin{table*}[t!]
	\begin{tabular}{| l | l | l | l |}
		\hline
		Sample & Thickness, \AA & Roughness, \AA   \\ \hline \hline
		SiO$_{2}$/Tb$_{18}$Co$_{82}$/Al$_2$O$_3$(18.8\AA)  & 309.5 & 7.6   \\ \hline
		SiO$_{2}$/Tb$_{24}$Co$_{76}$/Al$_2$O$_3$(30\AA)  & 261.4 & 5.2   \\ \hline
		SiO$_{2}$/Tb$_{24}$Co$_{76}$/Al$_2$O$_3$(39.5\AA)  & 378.5 & 6.5   \\ \hline
		SiO$_{2}$/Tb$_{26}$Co$_{74}$/Al$_2$O$_3$(39.7\AA)  & 359.7 & 5.5   \\ \hline
		SiO$_{2}$/Tb$_{29}$Co$_{71}$/Al$_2$O$_3$(31.2\AA)  & 260.7 & 5.3   \\ \hline
		SiO$_{2}$/Tb$_{30}$Co$_{70}$/Al$_2$O$_3$(35.6\AA)  & 344.5 & 5.5   \\ \hline 
		SiO$_{2}$/Al$_{80}$Zr$_{20}$(33\AA)/Tb$_{18.4}$Co$_{81.6}$/Al$_{80}$Zr$_{20}$(29\AA) & 165 & 8.6   \\  \hline
		SiO$_{2}$/Al$_{80}$Zr$_{20}$(31.6\AA)/Tb$_{22.4}$Co$_{77.6}$/Al$_{80}$Zr$_{20}$(33.7\AA) & 163.8 & 11.2   \\ \hline 
		SiO$_{2}$/Al$_{80}$Zr$_{20}$(38.6\AA)/Tb$_{22.4}$Co$_{77.6}$/Al$_{80}$Zr$_{20}$(27.2\AA) & 252.4 &   \\ \hline 
		SiO$_{2}$/Al$_2$O$_3$(13\AA)/Tb$_{18}$Co$_{82}$/Al$_2$O$_3$(30\AA) & 247 & 7.8  \\ \hline
		SiO$_{2}$/Al$_{2}$O$_{3}$(35.6\AA)/Tb$_{18.6}$Co$_{81.4}$/Al$_{2}$O$_{3}$(35\AA) & 165.6 & 6.1   \\ \hline
		SiO$_{2}$/Au(295.6\AA)/Tb$_{29}$Co$_{71}$/Al$_2$O$_3$(59.9\AA)  & 259.5 & 19.9   \\ \hline
		\hline
	\end{tabular}
	\caption{Summary of TbCo Layer thickness and roughness (second and third columns, respectively) as obtained from the \textit{GenX} fits in various samples of different Tb:Co compositions, as well as buffer and cap layers.}
	\label{tab:1}
\end{table*}

\begin{figure}[htp!]
	\centering
	\includegraphics[width=.9\columnwidth]{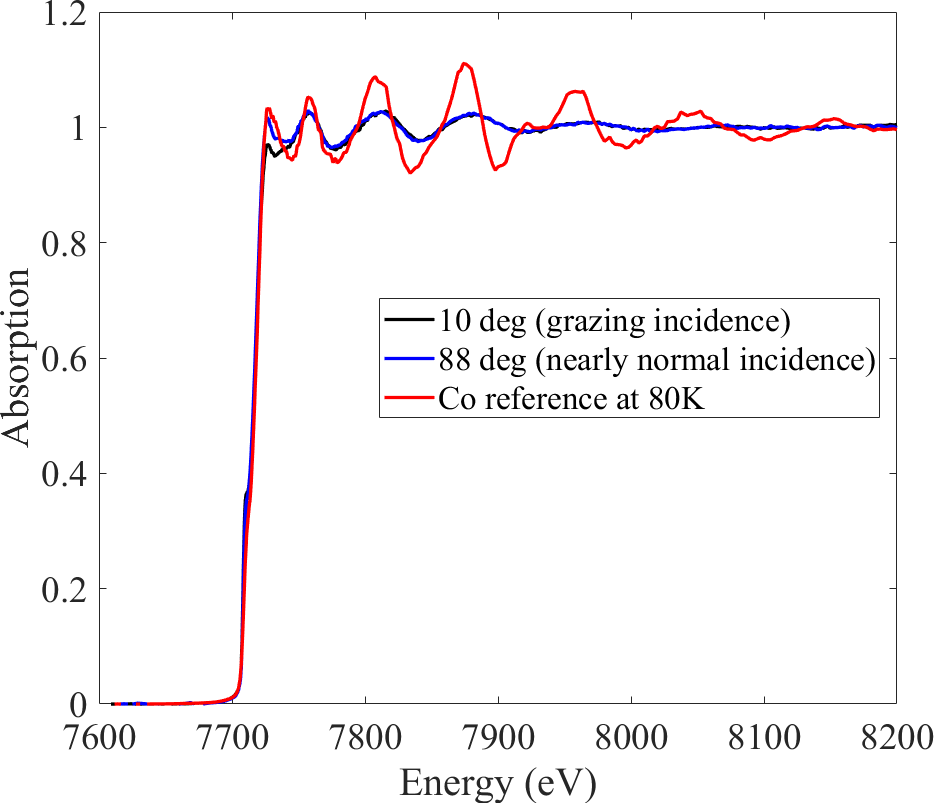}
	\caption{EXAFS spectra for $\mathrm{Tb_{18}Co_{82}}$ film, deposited onto a fused silica substrate and capped with AlZr, measured at two nearly orthogonal directions. Crystalline Co reference is shown by a red line.}
	\label{fig:EXAFS}
\end{figure}

Complementary to x-ray characterization, we performed EXAFS (SuperXAS - X10DA beamline at PSI, Switzerland) measurements to confirm the amorphous structure of the TbCo thin films. Figure \ref{fig:EXAFS} shows EXAFS absorption spectra for a sample with the following structure: fused silica/$\mathrm{Al_{80}Zr_{20}/Tb_{18}Co_{82}/Al_{80}Zr_{20}}$. The spectra were recorded at the Co-K edge and for two orthogonal directions: at a grazing incidence (10 deg from the sample plane, black curve) and at a nearly normal incidence (88 deg from the sample plane, blue curve). The absorption curve of the crystalline Co film was recorded at 80 K and is shown as a red reference curve. From the recorded spectra at grazing and a nearly-normal incidence, no significant  structural anisotropy can be observed.

\section{Magneto-optical measurements}

\begin{figure}[htb!]
	\centering
	\includegraphics[width=0.9\columnwidth]{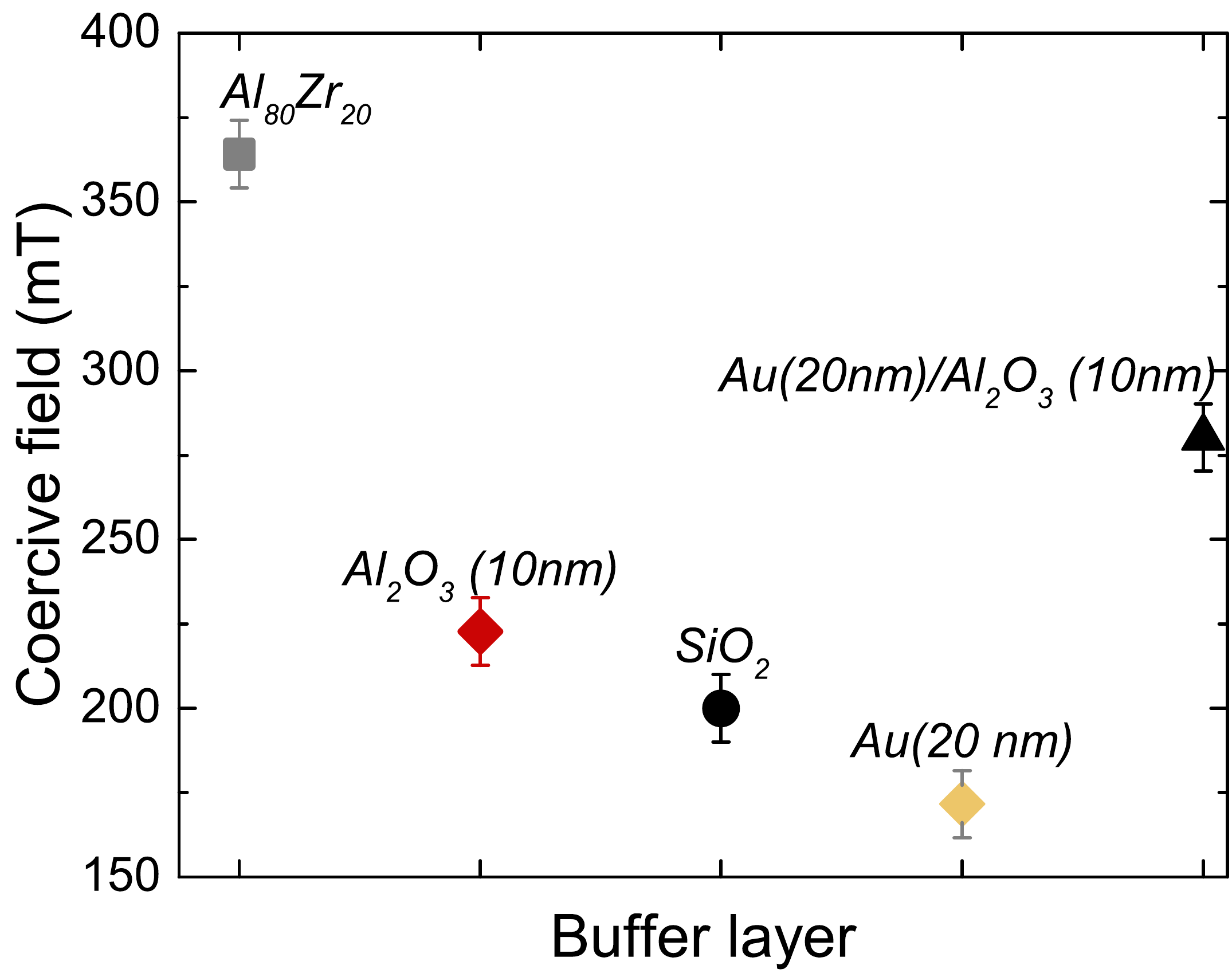}
	\caption{Coercive field dependence for 20 nm thick $\mathrm{Tb_{18}Co_{82}}$ films deposited on different buffer layers.}
	\label{fig:AlOx_buffer_dependence}
\end{figure}

We plot the coercive field dependence on the buffer layer for the 20 nm thick $\mathrm{Tb_{18}Co_{82}}$ films in the Fig. \ref{fig:AlOx_buffer_dependence}. It can be observed that the sample prepared onto Al$_{80}$Zr$_{20}$ buffer layers exhibits highest, while deposition directly onto Au buffer layer results in a significant reduction of coercive field.

\begin{figure*}
	\centering
	\includegraphics[width=0.85\linewidth]{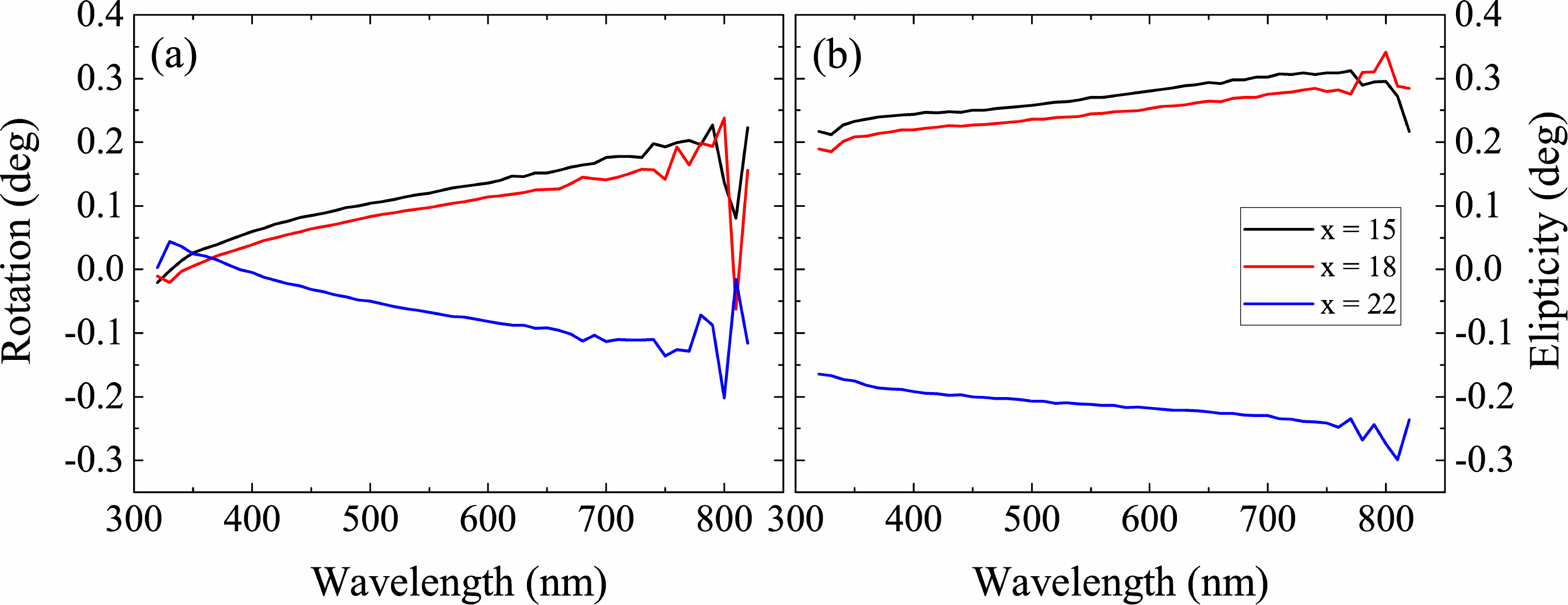}
	\caption{Dependence of Kerr rotation and ellipticity on the wavelength of incident light for 20 nm thick $\mathrm{Tb}_x\mathrm{Co}_{100-x}$ samples with x = 15, 18.6, 22.4 at.$\%$ Tb}
	\label{fig:El_and_rot_disp}
\end{figure*}

\begin{figure}
	\centering
	\includegraphics[width=0.9\columnwidth]{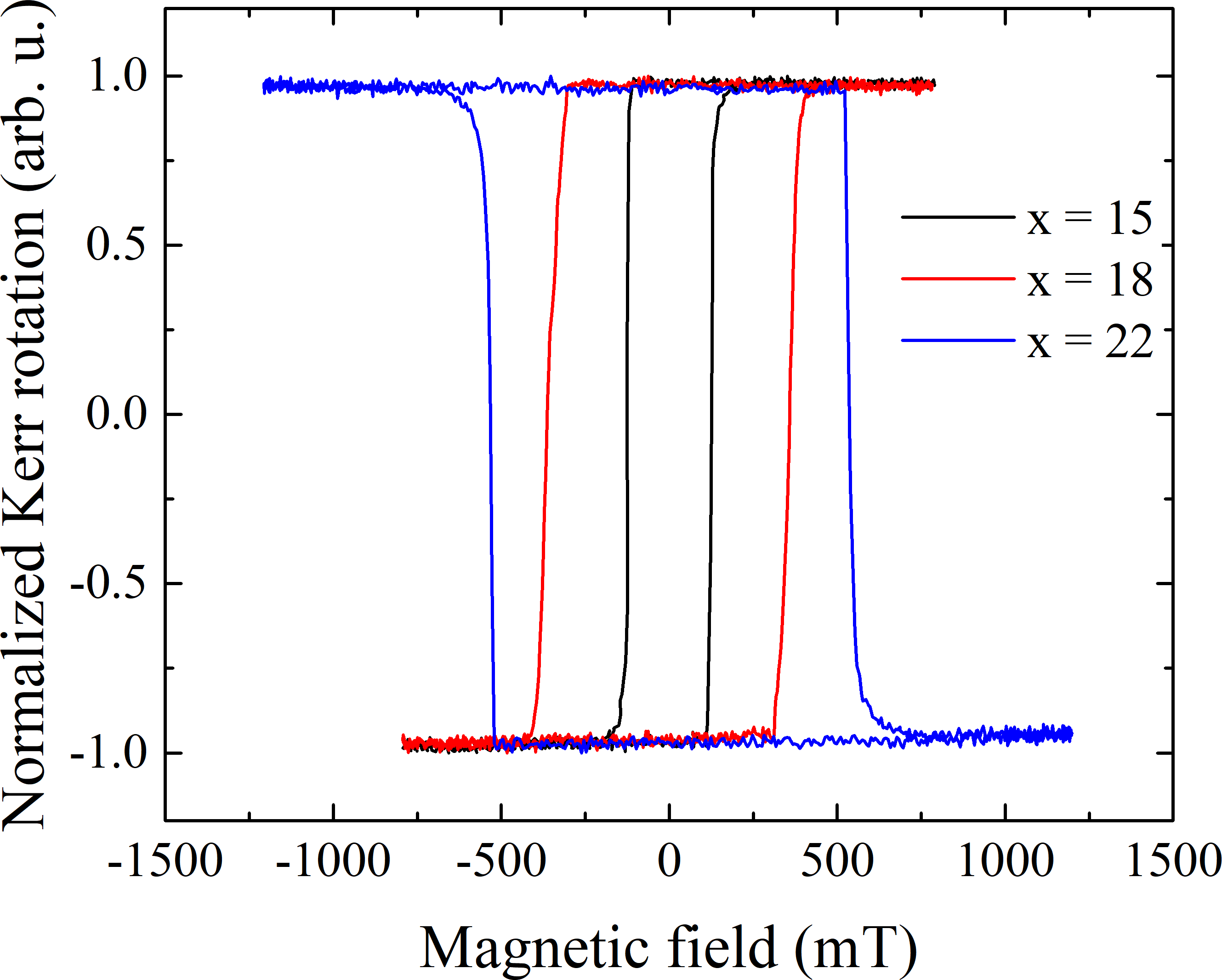}
	\caption{PMOKE measurements of $\mathrm{Al}_{80}\mathrm{Zr}_{20}/\mathrm{Tb}_x\mathrm{Co}_{100-x}/\mathrm{Al}_{80}$ $\mathrm{Zr}_{20}$ films, where x = 15, 18 and 22 at.$\%$. Loops were recorded using incident light with wavelength of 530nm.}
	\label{fig:AlZr_loops}
\end{figure}

Samples with 15, 18 and 22 at. $\%$ Tb prepared onto and capped with Al$_{80}$Zr$_{20}$ buffer layers, were measured employing polar magneto-optical effect (PMOKE). Magnetization loops were measured at room temperature using a PMOKE setup in reflection geometry as a function of wavelength of the incident light. The incident light wavelength was swept from 400 to 800 nm, while an out-of-plane static magnetic field of 900 mT was applied. Fig. \ref{fig:El_and_rot_disp} demonstrates the wavelength dependence of Kerr rotation and ellipticity. It can  be observed that in contrast to films with 15 and 18 at.$\%$ Tb, the measured out-of-plane hysteresis loop of the sample with 22 at.$\%$ Tb appears to have changed the sign/direction (See Fig. \ref{fig:AlZr_loops}), in agreement with the Kerr rotation and ellipticity spectra.

Since MOKE is mostly sensitive to the TM component, with the careful alignment of polarizer and analyzer in the MOKE setup one can determine the crossover from the TM to RE dominant composition via observation of the magnetization loop sign. When the alloy is TM dominated, the net magnetic moment follows the direction of the external magnetic field. When the alloy becomes RE- dominated, the total magnetic moment is antiparallel to an external magnetic field, and therefore for positive applied magnetic field the saturation is negative, and vice versa. This indicates that the compensation point for films prepared onto AlZr buffer layer is around 20 at.$\%$ Tb at room temperature, which is in agreement with results presented in a previous study by \citet{Frisk_2015}. 

\begin{figure}
	\centering
	\includegraphics[width=0.9\columnwidth]{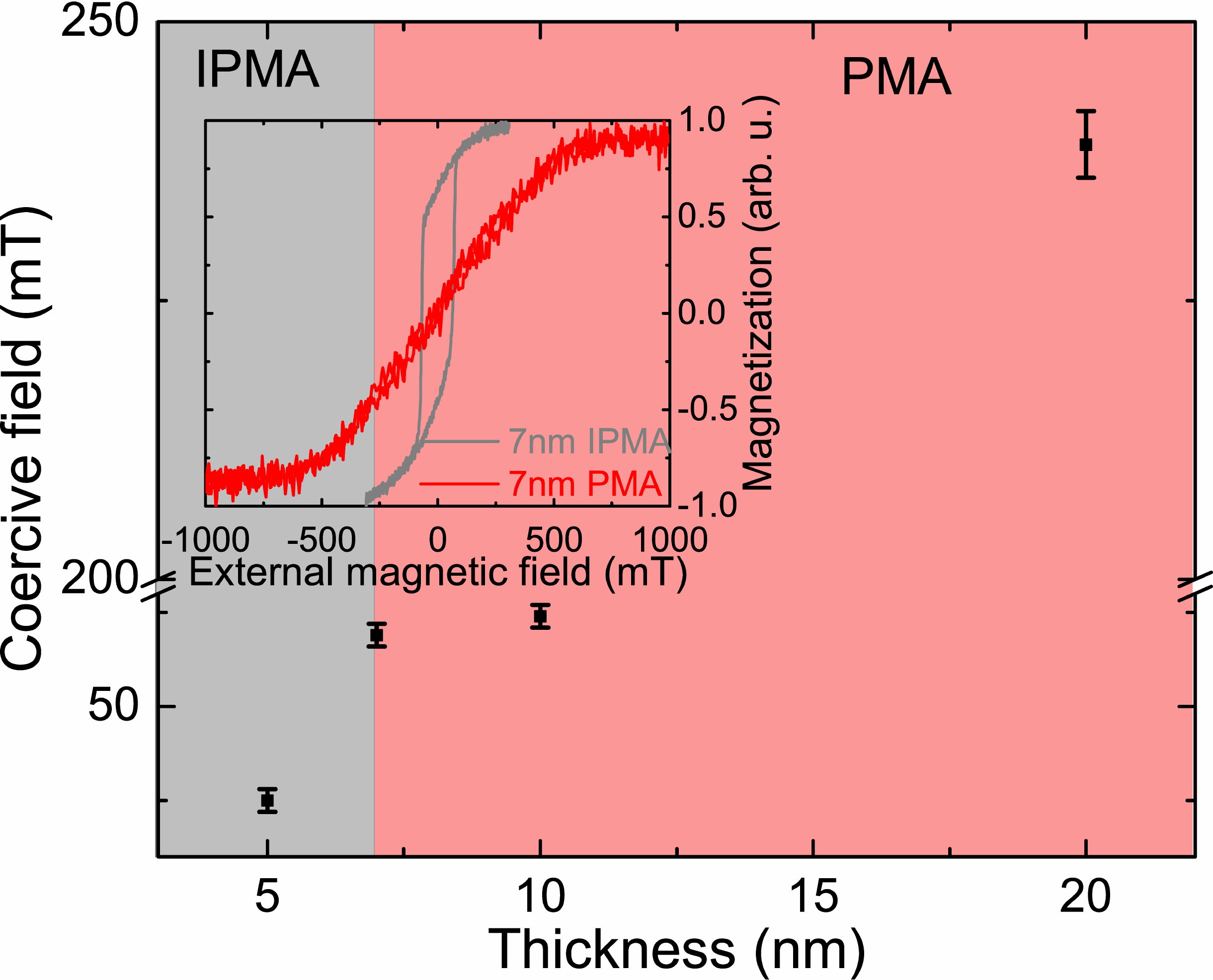}
	\caption{PMOKE measurements of $\mathrm{Tb_{18}{Co}_{82}/Al_2O_3}$ films of varying thickness. Loops were recorded using incident light with wavelength of 530nm. Inset shows MOKE loops recorded for a 7 nm thick sample in longitudinal and polar configurations (IPMA and PMA, respectively). }
	\label{fig:Hc_thickness}
\end{figure}

We measured the thickness dependence of the coercive field for $\mathrm{Tb_{18}Co_{82}/Al_2O_3}$ films using PMOKE setup with 530 nm wavelength of incident light (Fig. \ref{fig:Hc_thickness}). It can be seen that 7 nm thickness of films is a threshold thickness between in-plane and out-of-plane magnetization. Thicker samples exhibit square out-of-plane loops for the film composition of $\mathrm{Tb_{18}Co_{82}}$.

\section{Kerr microscopy}
\begin{figure}[ht!]
	\centering
	\includegraphics[width=0.9\columnwidth]{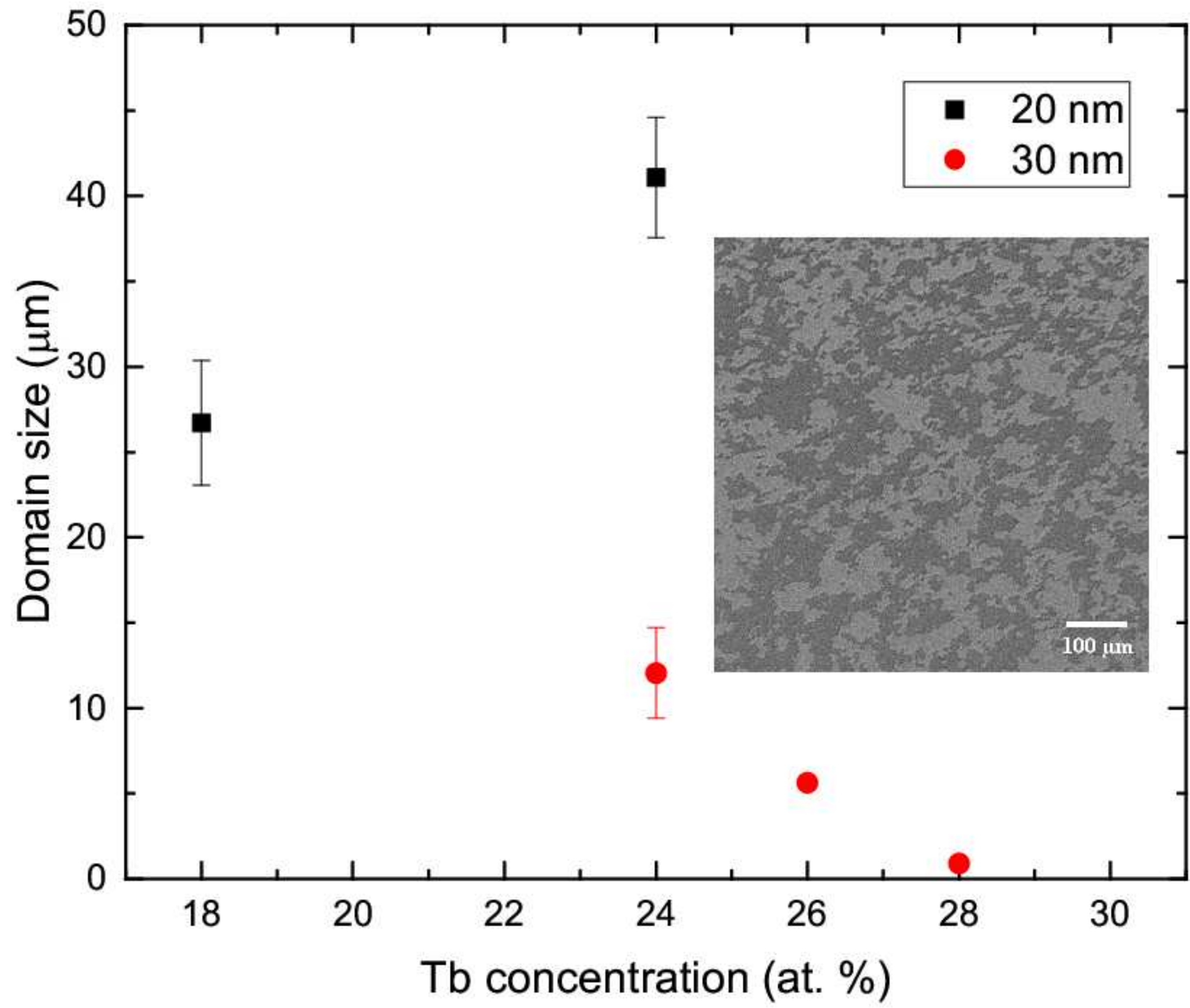}
	{\caption{Average magnetic domain size dependence on the Tb content in Tb$_{x}$Co$_{100-x}$ films for samples capped with  Al$_2$O$_3$ layers. Inset shows a Kerr microscopy image of the 20 nm thick Tb$_{24}$Co$_{76}$ sample.}}
	\label{Domains}
\end{figure}

We imaged the magnetic domain structure of our films using Kerr microscopy. The imaging was performed after demagnetizing sample in the oscillating magnetic field with an amplitude, decaying in time. This way the remanent magnetic state was observed. Here we present results of an average domain size variation for Tb-rich 30 nm thick films. The domain size computed by the pair correlation function (PCF) (see Fig. \ref{Domains}) is from tens to a few $\mu m$ in size for all of the films, and therefore the domains are large enough in order for the films to be used for patterning of nanosized elements exhibiting a single domain state. Reduction in the average domain size with increasing Tb content can be observed, similarly to the coercive field, which decreases with increasing Tb content. Equilibrium domain size was used as a criterion for the observation of the AOS in TbCo alloys by \citet{hadri_domain_size_criterion_PhysRevB.94.064419}, where the criterion is for the domain to be larger than the laser spot size. In our work, the laser spot size (FWHM), depending on the fluence, was in the range of $50 - 80 \mu m$, which is much larger than the observed domain size. Nonetheless, we observed AOS for a range of TbCo compositions. 


\section{Optical characterization}

We performed optical characterization by means of ellipsometry. We measured a range of Tb$_x$Co$_{100-x}$ films with different $x$, capped with Al$_2$O$_3$ in order to prevent oxidation of the film after deposition since all the characterization was not performed \textit{in situ}. The ellipsometric quantity $\rho$=$\frac{r_p}{r_s}$=$tan\psi\cdot e^{i\delta}$ \cite{tompkins_handbook_2005} was measured and used to extract the refractive index $n$ and extinction coefficient $k$ by means of \textit{GenX} fits \cite{GenX}. The \textit{GenX} software \cite{genx_source} is generally used for X-ray and neutron reflectivity data fitting, but by rescaling the X-ray scattering lengths (Eq. \ref{eq:Df}) one can implement visible light wavelengths. 
The complex refractive index and \textit{GenX} quantity scattering length density $f$ are related through the dielectric susceptibility $\chi$:

\begin{equation}
D \cdot f = -\chi \cdot \frac{\pi}{r_e \cdot \lambda^2},
\label{eq:Df}
\end{equation}

where $D$, $r_e$ and $\lambda$ are a material density, the electron radius and wavelength of incident light, respectively, and since $\epsilon=(n+ik)^2=1+\chi$, \\

\begin{equation} 
D\cdot f=-[(n+ik)^2-1] \cdot \frac{\pi}{r_e \cdot \lambda^2}
\end{equation}

The GenX-program calculates the complex wave field matrix W of the reflected wave, from which the ellipsometric data can be calculated and fitted.

\begin{equation}
\label{eq:Optical_matrix}
\left[ \begin{array}{c} E_s \\ E_p \end{array} \right]= 
\left[ \begin{array}{c} W_{00}  W_{01}\\ W_{10}   W_{11} \end{array} \right] \times \left[ \begin{array}{cc} E_{s0} \\ E_{p0} \end{array} \right]
\end{equation}

The script for ellipsometry data fitting can be found on the Zenodo repository \cite{zenodo}. 

After the extraction of $n$ and $k$ one can compute optical conductivity for the films: 

\begin{equation}
\sigma(\lambda)=\frac{nk\omega}{2\pi}=\frac{nkc}{\lambda}.
\end{equation}

\begin{figure}
	\centering
	\includegraphics[width=0.9\columnwidth]{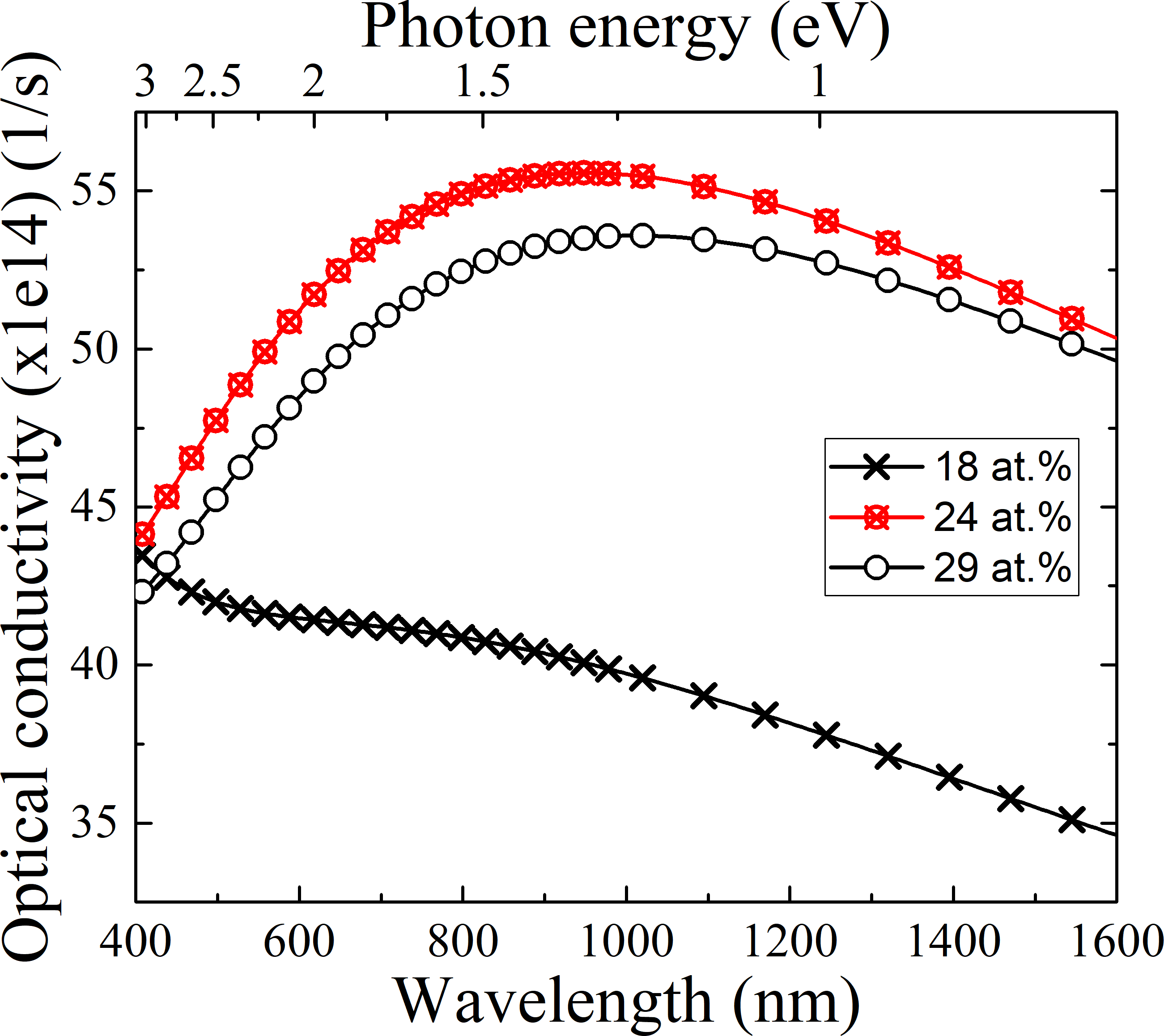}
	\caption{Dispersion of optical conductivity of Tb$_x$Co$_{100-x}$ films.}
	\label{fig:Optical_conductivity}
\end{figure}

The optical conductivity dispersion for our investigated Tb$_x$Co$_{100-x}$ films is shown in Fig. \ref{fig:Optical_conductivity}. The peak at around 900 nm wavelength can be observed for samples containing 24 and 29 at. $\%$ Tb, while sample with 18 at. $\%$ Tb appears to have a peak at shorter wavelengths. Optical conductivity dispersions can provide information about electronic transitions induced by optical excitations \cite{miller_optical_1974}. This information could be used for modelling electronic density of states and electronic transitions within the metallic compounds. However, this is out of scope of this work.

\begin{figure}
	\centering
	\includegraphics[width=0.9\columnwidth]{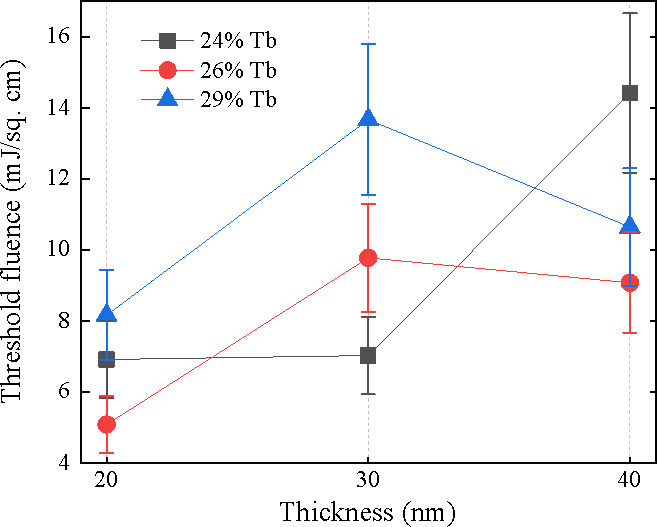}
	\caption{Dependence of threshold fluence on film thickness for three different compositions of  $\mathrm{Tb}_{x}\mathrm{Co}_{100-x}$}
	\label{fig:Fluence_Threshold_thickness_RU}
\end{figure}

\section{Laser pulse parameters for All-Optical Helicity Dependent Switching}

We plot the threshold fluence, above which AO-HDS/demagnetization can be observed, as a function of film thickness in Figure \ref{fig:Fluence_Threshold_thickness_RU}. While there is no significant variation observed overall within the experimental error bars, the threshold fluence appears to slightly increase with the film thickness. We speculate that increasing the film thickness leads to both better switching quality (as discussed in the main text) and an increasing fluence threshold due to the improved heat dissipation associated with thicker films.

\FloatBarrier

\end{document}